\documentclass[12pt]{article}
\usepackage{arxiv}
\usepackage{times}
\usepackage{hyperref}
\usepackage[pdftex]{graphicx}
\usepackage{subfig}
\usepackage{wrapfig}
\graphicspath{{Images_SprayCraft/}}
\usepackage{xcolor}
\usepackage[T1]{fontenc}
\usepackage{cite}
\usepackage{amsmath}
\usepackage[english]{babel}
\usepackage{multirow}
\usepackage{citesort}
\usepackage{mathptmx} 
\usepackage{xcolor}
\usepackage{rotating}
\usepackage{rotfloat}
\usepackage{amsmath}
\usepackage{amssymb}
\usepackage{amsfonts}
\usepackage{array}
\usepackage[ruled, linesnumbered, vlined]{algorithm2e}
\usepackage{gensymb}
\usepackage{textcomp}
\usepackage{epstopdf}
\usepackage{balance}
\usepackage{times}
\usepackage{url}
\renewcommand{\baselinestretch}{1}
\usepackage{graphicx}
\usepackage{booktabs}

\usepackage{balance}
\usepackage{fancyhdr}
\SetKwInOut{KwIn}{Input}
\SetKwInOut{KwOut}{Output}
\usepackage{authblk}
\usepackage[utf8]{inputenc} 
\usepackage{multicol} 
\usepackage{orcidlink}
\usepackage{hyperref}
\usepackage{tikz}

\title{SprayCraft: Graph-Based Route Optimization for Variable Rate Precision Spraying}
\date{} 
\author{}

\begin{document}
	
	\maketitle
	\vspace{-2.5cm}
	\begin{multicols}{3}
		\centering
		\textbf{Kiran K. Kethineni}\orcidlink{0009-0004-6853-6749} \\
		Computer Science \& Eng. \\
		University of North Texas, USA \\
		kirankumar.kethineni@unt.edu
		
		\columnbreak
		
		\centering
		\textbf{Saraju P. Mohanty}\orcidlink{0000-0003-2959-6541} \\
		Computer Science \& Eng. \\
		University of North Texas, USA \\
		saraju.mohanty@unt.edu
		
		\columnbreak
		
		\centering
		\textbf{Elias Kougianos}\orcidlink{0000-0002-1616-7628} \\
		Electrical Eng. \\
		University of North Texas, USA \\
		elias.kougianos@unt.edu
	\end{multicols}
	
	\begin{multicols}{2}
		\centering
		\textbf{Sanjukta Bhowmick}\orcidlink{0000-0001-8550-5371} \\
		Computer Science \& Eng. \\
		University of North Texas, USA \\
		Sanjukta.Bhowmick@unt.edu
		
		\columnbreak
		
		\centering
		\textbf{Laavanya Rachakonda}\orcidlink{0000-0002-7089-9029} \\
		Computer Science \\
		University of North Carolina Wilmington, USA \\
		rachakondal@uncw.edu

	\end{multicols}
	
	\vspace{10pt}

	\pagestyle{fancy}
	\fancyhead{}
	\fancyhead[L]{SprayCraft: Graph-Based Route Optimization for Variable Rate Precision Spraying}

	\begin{abstract}

	To efficiently manage plant diseases, Agriculture Cyber-Physical Systems (A-CPS) have been developed to detect and localize disease infestations by integrating the Internet of Agro-Things (IoAT). By the nature of plant and pathogen interactions, the spread of a disease appears as a focus with density of infected plants and intensity of infection diminishing outwards. This gradient of infection needs variable rate and precision pesticide spraying to efficiently utilize resources and effectively handle the diseases. This article, SprayCraft presents a graph based method for disease management A-CPS to identify disease hotspots and compute near optimal path for a spraying drone to perform variable rate precision spraying. It uses graph to represent the diseased locations and their spatial relation, Message Passing is performed over the graph to compute the probability of a location to be a disease hotspot. These probabilities also serve as disease intensity measures and are used for variable rate spraying at each location. Whereas, the graph is utilized to compute tour path by considering it as Traveling Salesman Problem (TSP) for precision spraying by the drone. Proposed method has been validated on synthetic data of locations of diseased locations in a farmland.

	
	\end{abstract}
			
	\section{Introduction}
	\label{sec:Introduction}

	The rapid increase in population and urbanization has resulted in environmental and climatic conditions that are unfavorable to agriculture posing threat to the global food security \cite{chakraborty2011climate}. In addition, due to the loss of biodiversity, pollution, and pathogen evolution, plant diseases are increasing and degrading food safety and security \cite{ristaino2021persistent} \cite{strange2005plant}. Since agriculture is the primary source of food for humanity and contributes significantly to farmers' income, the impact of plant diseases is especially concerning. These diseases are claiming about 20\% of the produce \cite{walker1983crop}, while the global population is growing at a rapid pace and is expected to reach 9.7 billion by 2050. As traditional agriculture struggles to meet rising demand, Internet-of-Agro-Things (IoAT) has been integrated into agriculture \cite{10026955} to develop Agriculture Cyber-Physical Systems (A-CPS) for smart and precision agriculture \cite{s19173796}.
	
	\begin{figure}[htbp]
		\centering
		\includegraphics[width=0.4\textwidth]{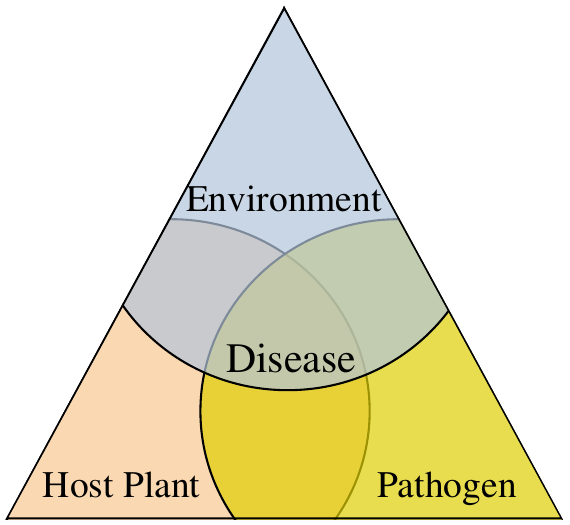}
		\caption{Disease triangle.}
		\label{fig:Disease_triangle}
	\end{figure}
	
	Plant diseases are caused by pathogens like bacteria, fungi, virus, nematodes and are transmitted by agents like insects, wind, water, physical contact \cite{singh2020transmission}  but get infested only if the environmental are favorable to the disease. This relation between host, pathogens and environment is represented in the Fig \ref{fig:Disease_triangle}. Since manual scouting is not feasible in case of large scale farming and humans also spread the diseases across the farm \cite{ranawaka2020homo}, many smart soil monitoring systems \cite{Rachakonda2024} and disease detection systems \cite{10561056} have been developed using IoT, imaging techniques and computer vision to identify disease infestations in farmland.

\begin{figure}[htbp]
		\centering
		\includegraphics[width=0.55\textwidth]{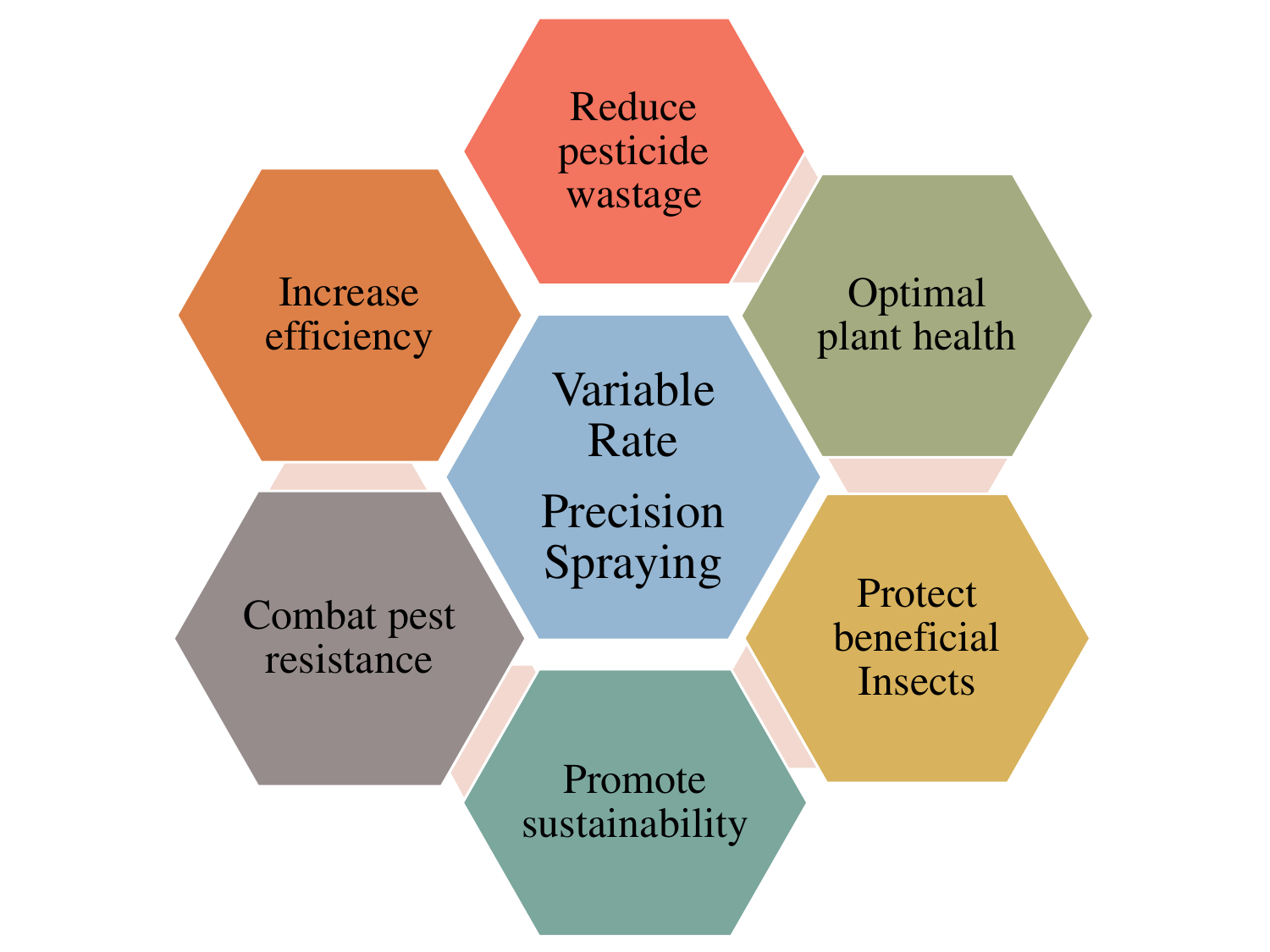}
		\caption{Advantages of Variable Rate Precision Spraying.}
		\label{fig:Variable_Rate_Precision_Spraying}
	\end{figure}

	In order to handle the disease spread, suppress the growth of the disease, farmland is sprayed with pesticides. But, when these pesticides are used in large amounts, they affect the environment \cite{Mahmood2016}, induce resistance in the crops and also affect the health of the humans who consume the produce \cite{toxics10060335}. So, pesticides should be precisely sprayed only on the affected areas to efficiently utilize them and prevent the negative effects of over-usage. Additionally, the severity of disease infestation varies across the farmland; areas that were initially affected or served as sources of the disease will generally have greater severity. The method of spraying only the diseased locations  and with dosages relative to the severity of infestation the location  is known as Variable Rate Precision Spraying and has various advantages as shown in Fig \ref{fig:Variable_Rate_Precision_Spraying}.

\begin{figure*}[htbp]
		\centering
		\includegraphics[width=0.8\textwidth]{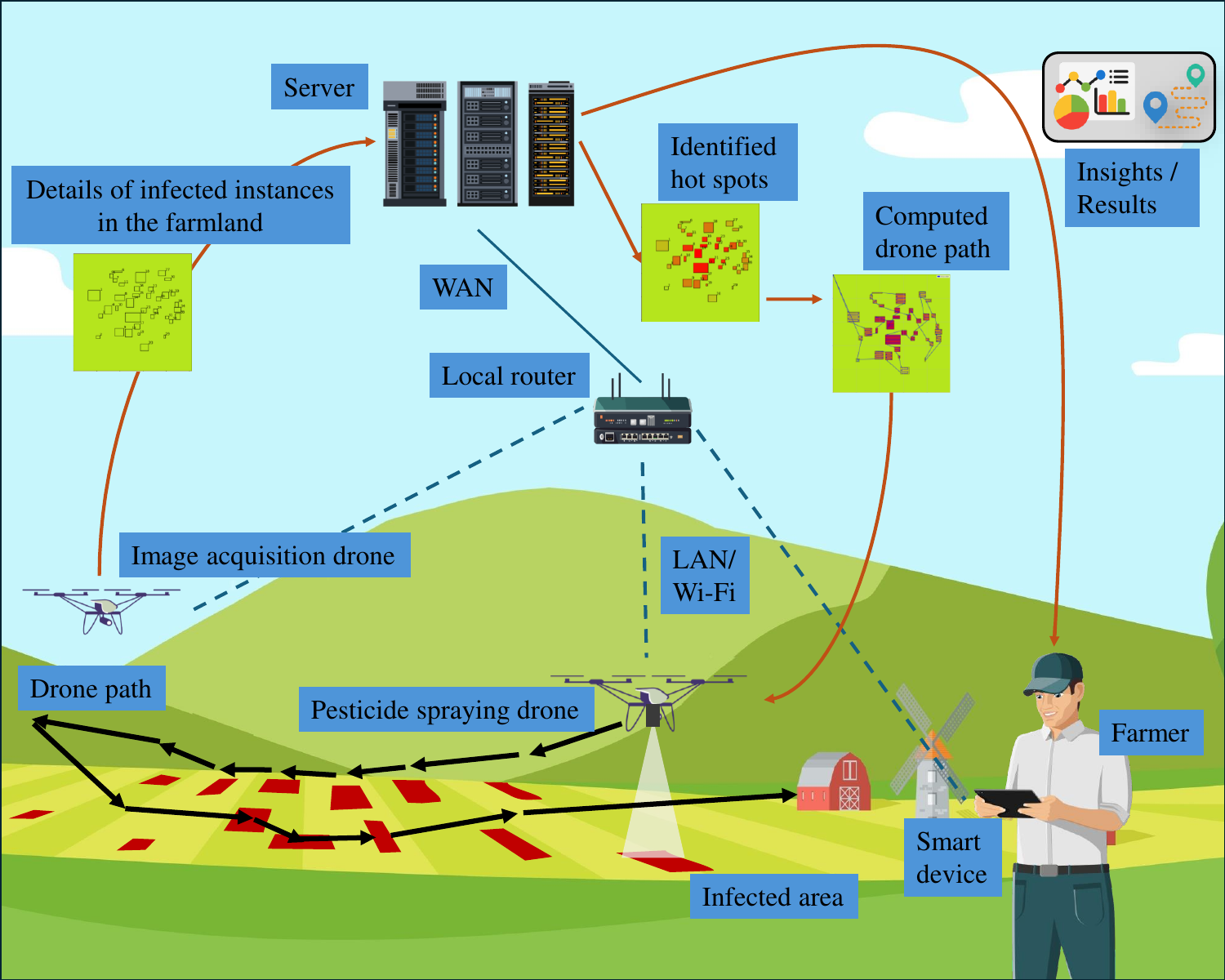}
		\caption{Overview of a Disease Management ACPS with SprayCraft.}
		\label{fig:SprayCraft_Overview}
	\end{figure*}

	Recently, as part of smart agriculture, drone based spraying systems \cite{MOGILI2018502} have been developed to reduce manual labor and automate the spraying of pesticides. On the same lines, we propose "SprayCraft" to compute near optimal route for variable rate precision pesticide spraying by drones in disease management ACPS as depicted by Fig \ref{fig:SprayCraft_Overview}.

	The article is organized as follows: Section \ref{sec:Novel_Contribution} presents the proposed solution to the defined problem and highlights its novelty. Section \ref{sec:Related_Prior_Works} reviews prior research related to the topic. Section \ref{sec:Graphs_for_Spatial_Analysis} and \ref{sec:Graphs_for_Routing} introduces the concept of graphs and their utility in spatial analysis and route optimization respectively paving the way for the graph-based hotspot identification proposed in Section \ref{sec:Hotspot_Detection} and route computation in Section \ref{sec:Route_Computation} . The proposed method is experimentally validated in Section \ref{sec:Experimental_Verification}, followed by discussion of the results in Section \ref{sec:Comparative_Perspective_With_Related_Works}. Section \ref{sec:Conclusion} concludes the article with remarks and future directions.
	
	\section{Novel Contributions of the Current Paper}
	\label{sec:Novel_Contribution}

	\subsection{The Problem Statement}
	\label{sec:The_Problem_Statement}
	
\begin{figure}[htbp]
		\centering
		\includegraphics[width=0.6\textwidth]{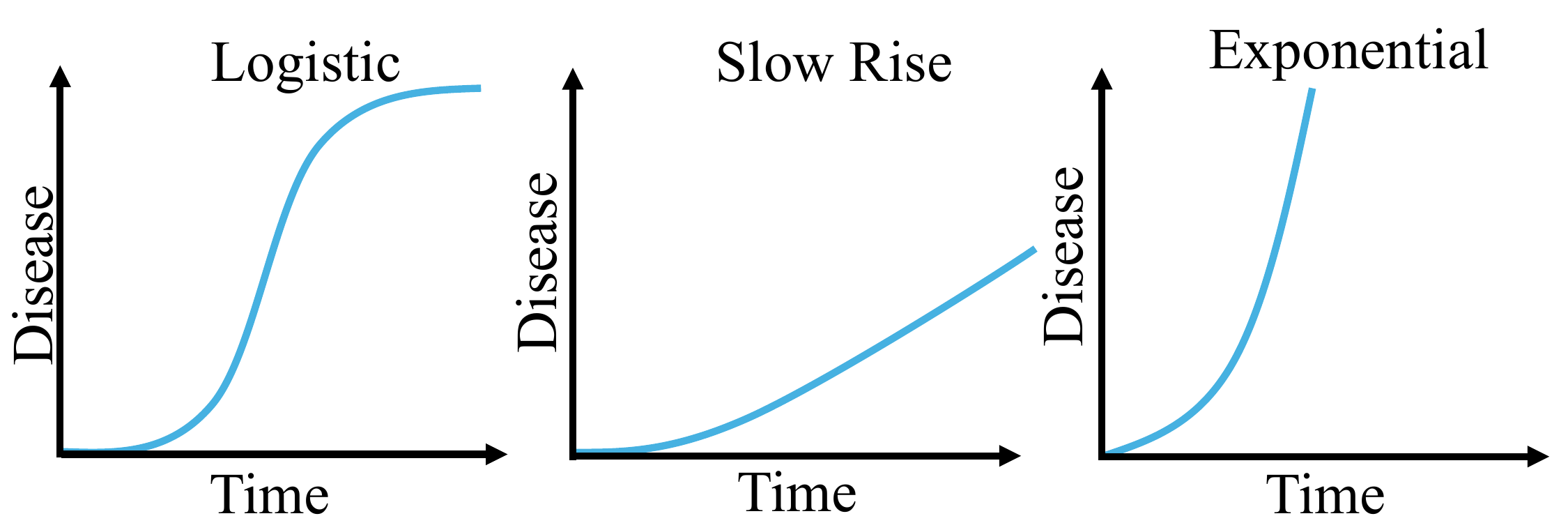}
		\caption{Disease Propagation Scenarios.}
		\label{fig:Disease_Propagation}
	\end{figure}

	Most of the diseases apart from seed born originate at a location and spread across due to pathogen interactions and environmental factors. When the spread of diseases is plotted with time on X axis and disease intensity on Y axis, the spread is categorized to three types. Logistic Growth: The disease would initially grow slow rate and advance with more number of instances being infected and accelerating the spread. When the disease is spread to most of the farmland, due to lack of resources the rate of spread would decrease and plateau. Slow Rise: The disease would transmit at slower pace and intensity increases steadily. Exponential Growth: This is the common pattern present in spread of most of the diseases. Gere, the rate of spread starts at slow rate and increases rapidly with time. In all the scenarios, the graph shows an upward trend as shown in Fig \ref{fig:Disease_Propagation} \cite{real1996spatial} indicating that the intensity of disease increases with time.
	
\begin{figure}[htbp]
		\centering
		\includegraphics[width=0.4\textwidth]{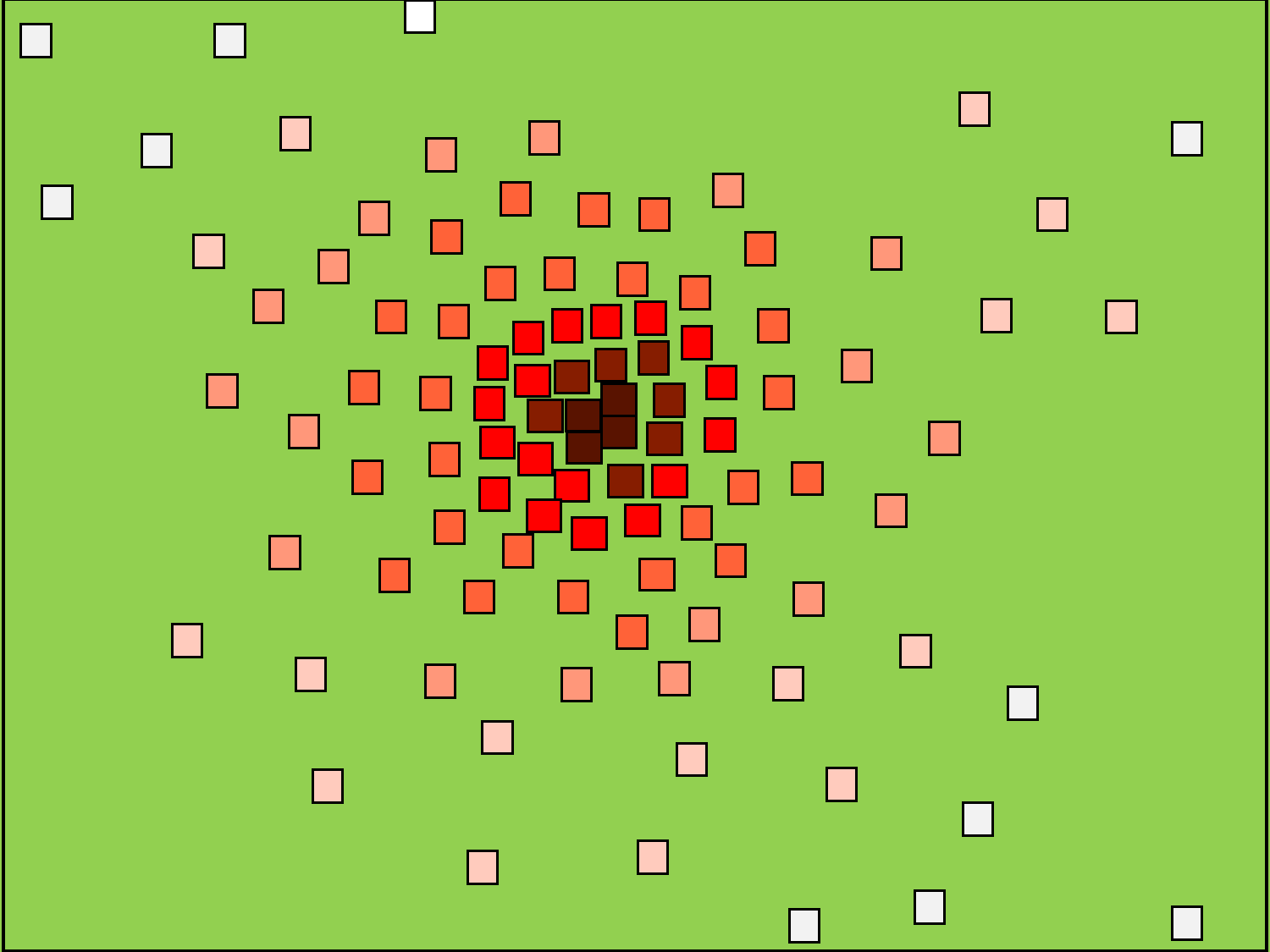}
		\caption{Representation of Hotspots in Disease Propagation.}
		\label{fig:Disease_Propagation_Hotspots}
	\end{figure}

	As the disease intensity increases with time, the density of the disease also increases radially with time and will have rougly circular pattern with a focus at the center \cite{arneson2001plant}. In addition, in case of farming involving very large area, environmental conditions are not same across the farm due to factors like elevation, composition of soil, ability of the soil to retain the soil \cite{plantegenest2007landscape}. As a result, parts of the farmland may have conditions that are favorable for diseases and act as source for diseases. The spatial area where the density of infestation, risk of transmission is higher, or the probability of that area being the source for the disease as shown in Fig \ref{fig:Disease_Propagation_Hotspots} is termed as a disease hotspot. So, to better handle the disease propagation, the pesticide dosage has to be proportional to the probability of that instance being a hotspot.

\begin{figure}[htbp]
		\centering
		\includegraphics[width=0.6\textwidth]{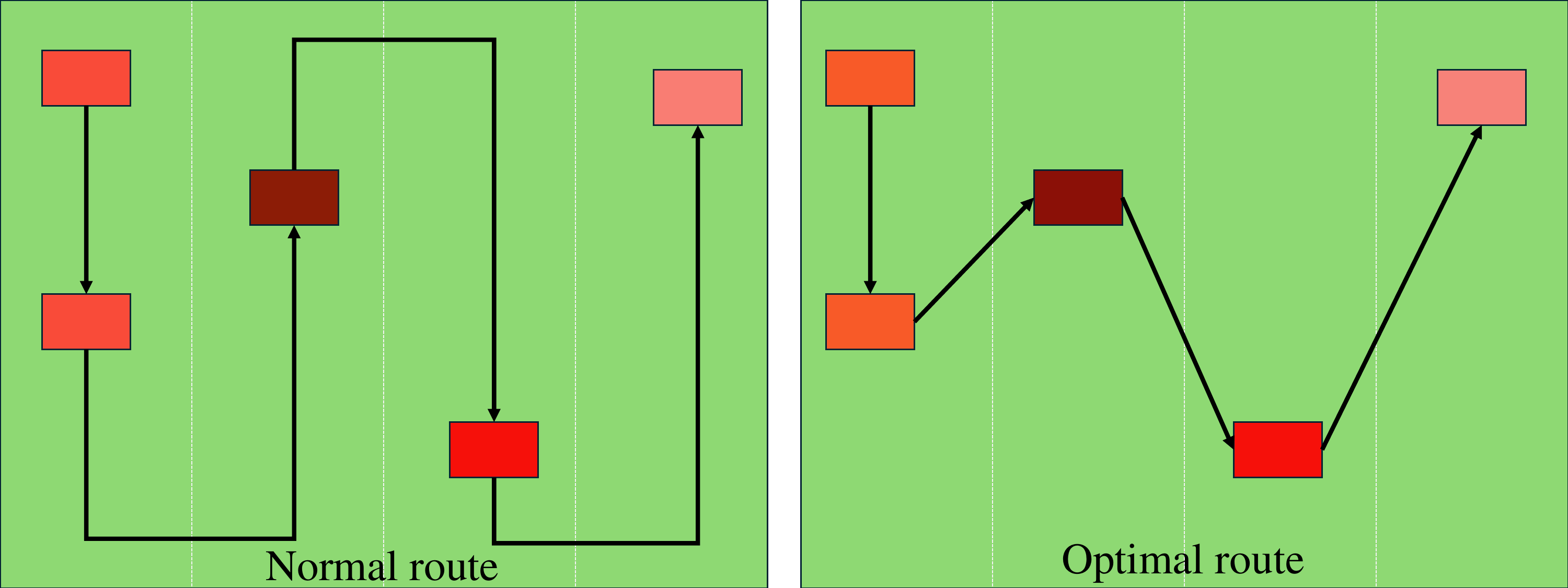}
		\caption{Different Paths for Drone Routing.}
		\label{fig:Optimal_Path}
	\end{figure}

	In addition to performing variable rate pesticide spraying, precision methods to reduce the usage of resources are also necessary to achieve sustainable agriculture \cite{Sahni2024}. There can be many ways to route a drone assigned with pesticide spraying across the diseased instances, as represented in Fig \ref{fig:Optimal_Path}. Therefore, the system must compute a route that minimizes travel time while maximizing coverage and the effectiveness of the pesticide relative to the degree of infestation at each location.

	\subsection{Proposed Solution of the Current Paper}
	\label{sec:Proposed_Solution}

	Based on the previous discussion and the interrelation between spatial patterns and disease spread as described in articles \cite{nelson2008detecting} and \cite{ristaino2000new}, it can be deduced that likelihood of  a plant being infected is influenced by its location and the health of plants nearby. As the disease progresses, the number of diseased instances surrounding an diseased instance also increases proportionally. This dynamic relationship contributes to a distinct disease spread pattern, wherein the source of the disease, or hotspots, exhibit the highest density of diseased instances. These hostpots can be identified by analyzing the neighboring instances and the spatial relation between them.
	
	The current article, "SprayCraft," which extends our previous work presented at the "2023 OITS International Conference on Information Technology (OCIT)" \cite{10467103}, presents a sophisticated graph-based  methodology designed to analyze the spatial relationships among diseased instances within farmland for the identification of disease hotspots. By employing message passing algorithms, the method evaluates the probability of each diseased instance being a hotspot. This probabilistic assessment aims to determine the relative pesticide dosage required for each diseased instance, optimizing the pesticide application process.
	
	These findings are instrumental in leveraging pesticide spraying drones \cite{drones6120383}, to implement variable rate spraying. Some spraying systems are equipped with nozzles that maintain a constant flow rate \cite{8239330}, whereas some possess the capability to modulate flow rates in real-time \cite{app8122482}. The proposed method considers these capabilities and computing routes for the spraying drone in each instance, depending on the specific type of spraying system employed.
	
	Using the earlier constructed graph, the algorithm seeks to identify a near-optimal route that connects all the diseased instances, framing the problem as a Traveling Salesman Problem (TSP). This approach ensures that the drone covers all affected areas efficiently, thereby enhancing the precision and effectiveness of pesticide application.
		
	\begin{figure*}[htbp]
		\centering
		\includegraphics[width=0.9\textwidth]{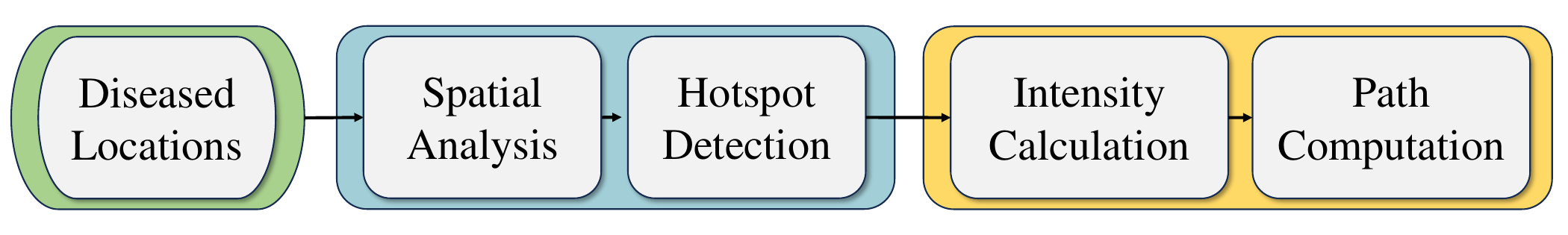}
		\caption{Block Diagram of the Proposed SprayCraft Method.}
		\label{fig:Block_diagram}
	\end{figure*}

	\subsection{Novelty and Significance of the Proposed Solution}
	
	\label{sec:Novel_Significance}
	
	The proposed method exhibits several elements of novelty, outlined as follows:
	\begin{enumerate}		
		
		\item Graph-Based Spatial Analysis for Disease Hotspot Identification: The article introduces a novel graph-based methodology to analyze spatial relationship between diseased instances. By using message passing algorithms, it accurately identifies disease hotspots and assists in understanding the disease spread in smart agriculture.
		
		\item Probabilistic Assessment for Variable Rate Spraying: The proposed method performs a probabilistic assessment of each diseased instance to determine if it is a hotspot. This assessment helps in the determination of relative pesticide dosage, enabling variable rate spraying. This approach optimizes pesticide usage, enhances the efficiency of disease management.
		
		\item Integration of Spraying Drone Capabilities: The method accounts for the varying capabilities of spraying systems, including both constant and variable flow rate nozzles. By computing customized routes for spraying drones based on the type of system used, the article offers a versatile solution that adapts to different technological setups, ensuring optimal pesticide application.
		
		\item Application of TSP for Optimal Spraying Routes: By using the Traveling Salesman Problem (TSP) framework, the article develops a near-optimal route for connecting all diseased instances. This application of TSP ensures that the spraying drone covers the affected areas while minimizing travel time and maximizing resource usage.
		
	\end{enumerate}

	\section{Related Prior Works}
	\label{sec:Related_Prior_Works}

	The pesticide spraying systems in agriculture evolved with advancements in IoT and integration of unmanned aerial vehicles (UAVs) and routing algorithms. The research in \cite{drones4030058} focused on optimizing UAV routes for precision pesticide spraying by identifying stressed crop regions, determining optimal spray points while minimize pesticide usage and flight time. Similarly, route planning for a spraying helicopter that needs to cover multiple areas was addressed in \cite{f12121658}. The study discussed the issues in creating routes for spraying pesticides within each area and the routes for traveling between different zones, but it did not cover the sprayer technology itself. A method similar to the proposed Spray-Craft is presented in \cite{plessen2024pathplanningspotspraying}, which optimizes the drone's path by using the Traveling Salesman Problem (TSP) for the global route and a headland path for the local route. In addition, \cite{plessen2024pathplanningspotspraying} also incorporated obstacle avoidance. In contrast to traditional TSP solutions, the authors of \cite{huang2023automatic} demonstrated the use of reinforcement learning to develop dynamic, environment-specific solutions. The proposed method considers factors such the location of infestations, density of target crops, slope and elevation of the surface to compute an efficient path for the UAV.

	Development of variable rate spraying system using PID and PWM control enabled the for UAV spraying systems to perform variable rate spraying as per prescription map \cite{app8122482}. In the article \cite{TEWARI202021}, a pesticide spraying prototype is presented that leverages computer vision techniques to assess the health of crops in its path in real-time, computes the disease severity, and applies pesticide accordingly, optimizing the treatment based on the severity of the disease. On the other hand, authors of \cite{7943794, 8875683} proposed route optimization to pay attention to specific regions in the field which can be also used to perform variable rate spraying to enhance the disease management and optimize resource utilization. 
	
	Further, efficient pesticide spraying coverage paths and task allocation among multiple UAVs has been proposed in \cite{9904967}. It formulated the problem as a constrained multiple traveling salesman problem, considered various constraints like power, pesticide availability while finding optimal route between instances. Authors of \cite{CONESAMUNOZ2016204} presented methods to consider more constraints for route optimization but lacked variable rate application capabilities or hotspot detection. Methods for determining the optimal route for each drone, taking into account the amount of pesticide to be applied at each location for variable-rate precision spraying, have been presented in \cite{9994402, 8264538}. Further, routing and coordination between ground vehicles and multiple UAVs have been optimized to enhance efficiency and coverage in pesticide application \cite{XU2024142429}. 
	
	While the models mentioned may have variable rate spraying or precision spraying capabilities, they do not include spatial analysis for hotspot detection. The proposed "SprayCraft" addresses this gap by performing spatial analysis to assess the probability of each node being a hotspot for disease. It calculates the required dosage based on this analysis and generates a near-optimal route for variable rate precision spraying.

	\begin{table*}[!h]
		\centering
		\caption{A brief summary of relevant literature.}
		\label{tab:comparison}
		\begin{tabular}{p{3cm}p{1cm}p{3.3cm}p{3.3cm}p{3.3cm}}
			\hline
			\textbf{Research} & \textbf{Year}  &  \textbf{Methodology} & \textbf{Optimization Goals} & \textbf{Remarks} \\
			\hline
			\hline
			Wen et al. \cite{app8122482} & 2018  & Variable rate spraying with PID control, PWM control & Implement variable rate spraying as per prescription map & Lacks spatial analysis and hotspot \newline detection\\
			\hline
			Plessen \cite{plessen2024pathplanningspotspraying} & 2024 & Global path optimization by TSP and local path optimization by headland path & Optimize area coverage path for spot spraying & Focuses on routing optimization but not variable rate spraying\\
			\hline
			Huang et al. \cite{huang2023automatic} & 2023 & Reinforcement learning for environment-specific solutions & Optimize spraying path to reduce use of pesticide and battery. & Focuses on routing optimization but not variable rate spraying\\
			\hline
			Tewari et al. \cite{TEWARI202021} & 2020 & Computer vision to estimate severity and perform variable rate spraying & Efficient pesticide usage with variable rate per instance. & Proposed methods for rover sprayer, route optimization is out of scope\\
			\hline
			Srivastava et al. \cite{drones4030058} & 2020 & Convex hull boundary construction and voronoi regions & Optimize UAV routes for precision spraying & Focuses on routing optimization but not variable rate spraying\\
			\hline
			Fang et al. \cite{f12121658} & 2021 & Hierarchical route optimization, variable application system & Minimize the dispatch routes and spraying routes & Does not cover intensity based rate calculation\\
			\hline
			Xu et al. \cite{9904967},\newline {Conesa-Muñoz} et al. \cite{CONESAMUNOZ2016204} & 2023, 2016 & Multi-UAS optimization algorithm for coverage path planning & Efficient spraying paths with \newline resource constraints & Does not support variable rate spraying\\
			\hline
			Nolan et al. \cite{7943794}, \newline Muliawan et al. \cite{8875683}  & 2017, 2019  & Focus at high intensity region by routing closer \slash more times & Route optimization with regional attention & Need predefined prescription map, cannot compute by itself\\
			\hline
			Zheng et al. \cite{9994402}, \newline Lal et al. \cite{ 8264538}  & 2022, 2017 & Multiple drone routing solved by MTSP algorithms & Optimal routes for multiple drones with variable dosage levels per \newline instance & Does not identify posible disease hotspots\\
			\hline
			\textbf{SprayCraft} & 2024  & Spatial analysis for hotspot \newline detection and tour all instances by TSP algorithm & Optimize length of route and \newline pesticide usage with variable rate per instance & Identifies potential hotspots and adapts route as per the type of spraying system \\
			\hline
		\end{tabular}
	\end{table*}

	\section{Graphs for Spatial Analysis}
	\label{sec:Graphs_for_Spatial_Analysis}

\begin{figure}[!htbp]
		\centering
		\includegraphics[width=0.6\textwidth]{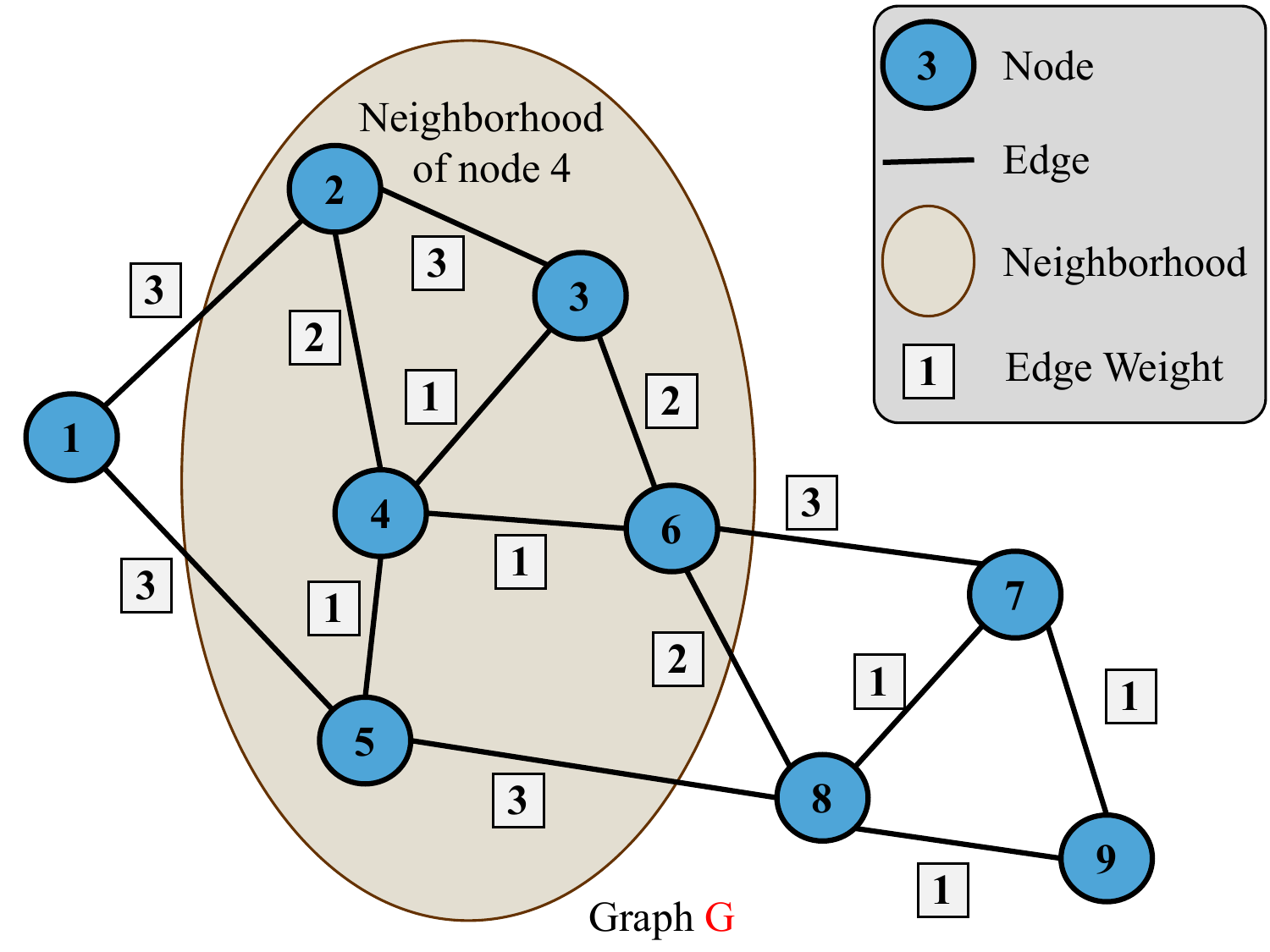}
		\caption{Example of a Graph.}
		\label{fig:Graph}
	\end{figure}

	A graph, as described by Newman \cite{newman2018networks}, is a data structure predominantly utilized to depict the relationships among a set of elements.
	A graph  is a data structure primarily used to represent the relation between a collection of elements. Any graph $G=(V,E)$ consists  of Nodes: $V = \{v_1,v_2,v_3, \ldots, v_n\}$  and Edges: $E = \{(v_i, v_j), (v_j, v_k), \ldots\}$  where nodes represent the elements and edges represent the connections between these nodes as depicted in Fig \ref{fig:Graph}. Weighted graphs have weights associated with each edge ($E = \{(v_i, v_j, w), (v_j, v_k, w'), \ldots\}$) while unweighted graphs do not have weights on their edges. In a graph, the neighborhood of a node $u \in V$, denoted as $N(u)$. consists of all the nodes that are directly connected to node $u$ by an edge.
	
	Graphs can be categorized based on their characteristics. Undirected graphs do not have any direction associated with their edges and express symmetry, while directed graphs have directed edges ($E = \{(v_i \rightarrow v_j), (v_j \rightarrow v_k), \ldots\}$) that represent one-way relationships in the direction of the edge. A graph is called a connected graph only if there is a path between every pair of nodes in the graph. Conversely, if there is no path between every pair of nodes, that graph is called as a disconnected graph.

\begin{figure}[!htbp]
		\centering
		\includegraphics[width=0.6\textwidth]{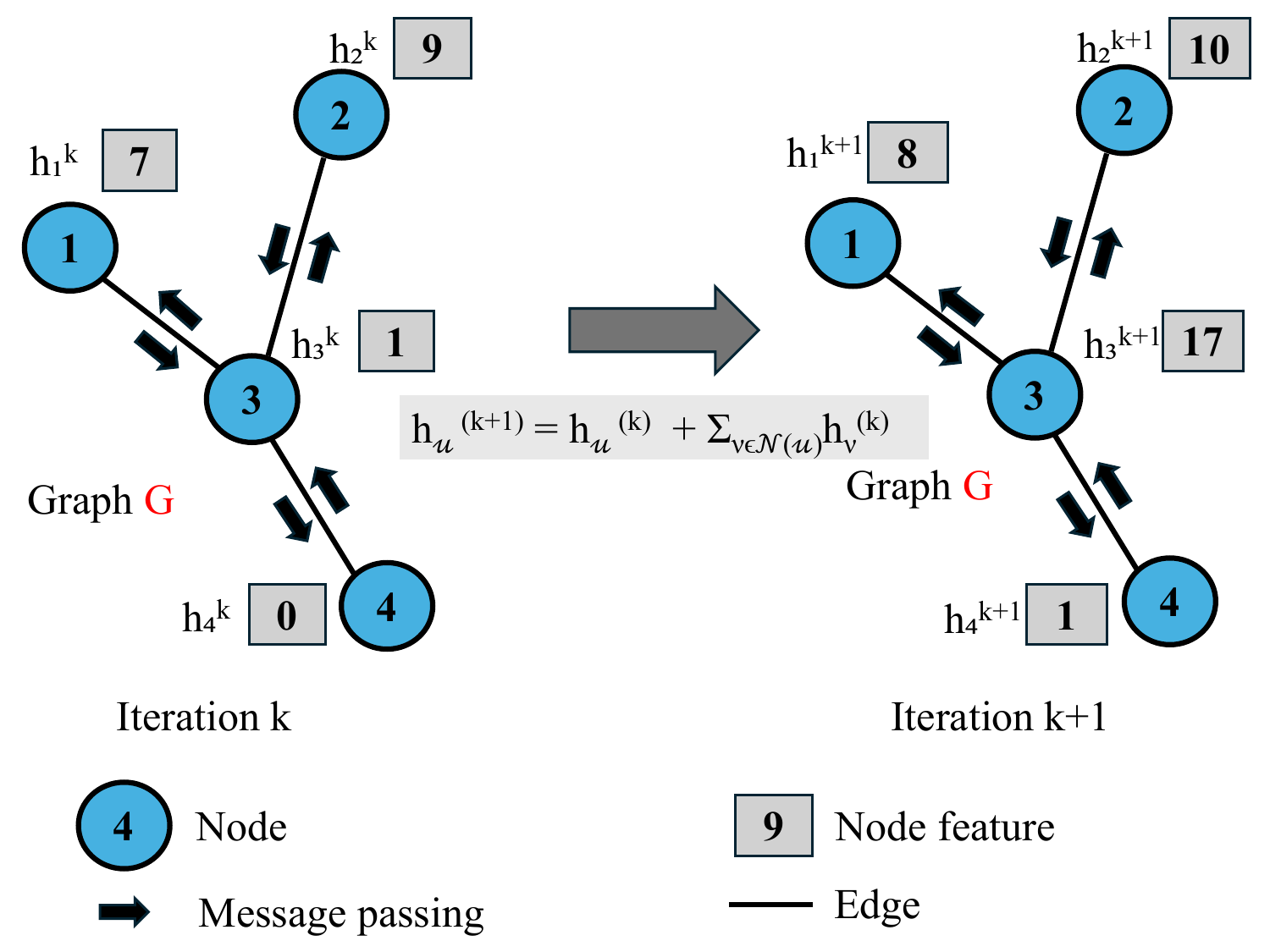}
		\caption{Illustration of Message Passing.}
		\label{fig:Message_Passing}
	\end{figure}

	Since graphs represent neighbors and their relationships, they can be crucial in analyzing how a node can influence its neighbors and how a node can be influenced by its neighbors. But, nodes alone do not sufficiently represent information or context. Given that spatial analysis seeks to explore the relationships and connectivity between nearby elements, representing the data as a graph $G = (V, E, F)$ with spatial attributes and characteristics as node features F = $\{f_1,f_2,f_3, \ldots, f_n\}$ can enable various methods for performing spatial analysis. Traditional neural network models are designed to work with data that has a feature set or with images, but they are not suited for graph data. Therefore, a new model called the Graph Neural Network (GNN) was developed to work with graph data. These models utilize message passing to update the feature vector of all nodes in the graph, thereby reflecting influence of all its neighbors, and perform machine learning tasks on the updated feature vectors. Message passing \cite{gilmer2017neural} is an aggregation method that involves propagating information between nodes to learn about their neighbors, as illustrated in Fig \ref{fig:Message_Passing}.

	The features learned from the neighbors are aggregated with the current features of node $u$ to update its feature vector from $h_u^k$ to a new feature vector $h_u^{k+1}$ as shown in the equation \ref{eq:Message_passing}.
	\begin{equation}
		\label{eq:Message_passing}
		h_u^{(k+1)} = \text{UPDATE}^{(k)} \left( h_u^{(k)},\text{AGG}^{(k)} \left( \{ h_v^{(k)} : v \in N(u) \} \right) \right),
	\end{equation}

	{
		Where: 
		\begin{align*}
		h^{(k+1)}_u & \text{ is the feature vector of node } u \text{ in } (k+1)^{th}\text{ iteration,} \\
		h^{(k)}_u & \text{ is feature vector of node } u \text{ in the } k^{th} \text{ iteration,} \\
		\text{{AGG}}^{(k)} & \text{ aggregates the feature vectors } h^{(k)}_v : v \in N(u) \text{ }\\
		 &   \text{ from the neighboring nodes } v \text{ in } N(u),\\
		\text{{UPDATE}}^{(k)} & \text{ intakes current feature vector } h^{(k)}_u \text{, aggregated}\\
		 &\text{messages to update feature vector to } h^{(k+1)}_u.
		\end{align*}
		}
	
	Message passing aids in learning about immediate neighbors and understanding the structure of the graph. Repeated message passing can diffuse features and information throughout the network, thereby revealing complex relationships. Due to the expressive power of graphs and message passing, plant disease location data is often represented as a graph to facilitate spatial analysis.
	
	\section{Graphs for Routing}
	\label{sec:Graphs_for_Routing}	
	
	Graph structures can also be instrumental in applications involving path computation. If a network of paths is represented as a graph where each intersection will be a node and the paths connecting will be edges, by analyzing the nodes and their neighbors, shortest path between two locations can be computed. For instance, consider a graph representing a water irrigation system where the junctions of pipelines or canals are nodes, and the segments of the pipelines or canals between the junctions are edges, with the capacity of the pipelines or canals as the edge weights. By running a maximum flow algorithm on this graph, we can gain insights into how much water is delivered to various parts of the farm during irrigation. By applying algorithms such as Dijkstra's or A*, we can determine the shortest or most efficient routes for irrigation pipelines. Graphs are especially useful for computing efficient routing paths for drones or automated vehicles that are supposed to cover \slash scout specific areas for tasks like irrigation, disease inspection, pesticide application while ensuring optimal resource utilization. Recent developments in Graph Neural Network (GNN) algorithms, such as Graph Attention Networks (GAT) and GraphSAGE, have enabled the increased use of graphs in real-time traffic forecasting and navigation applications, where graphs can be dynamic.
	
	\begin{figure}[!htbp]
		\centering
		\includegraphics[width=0.6\textwidth]{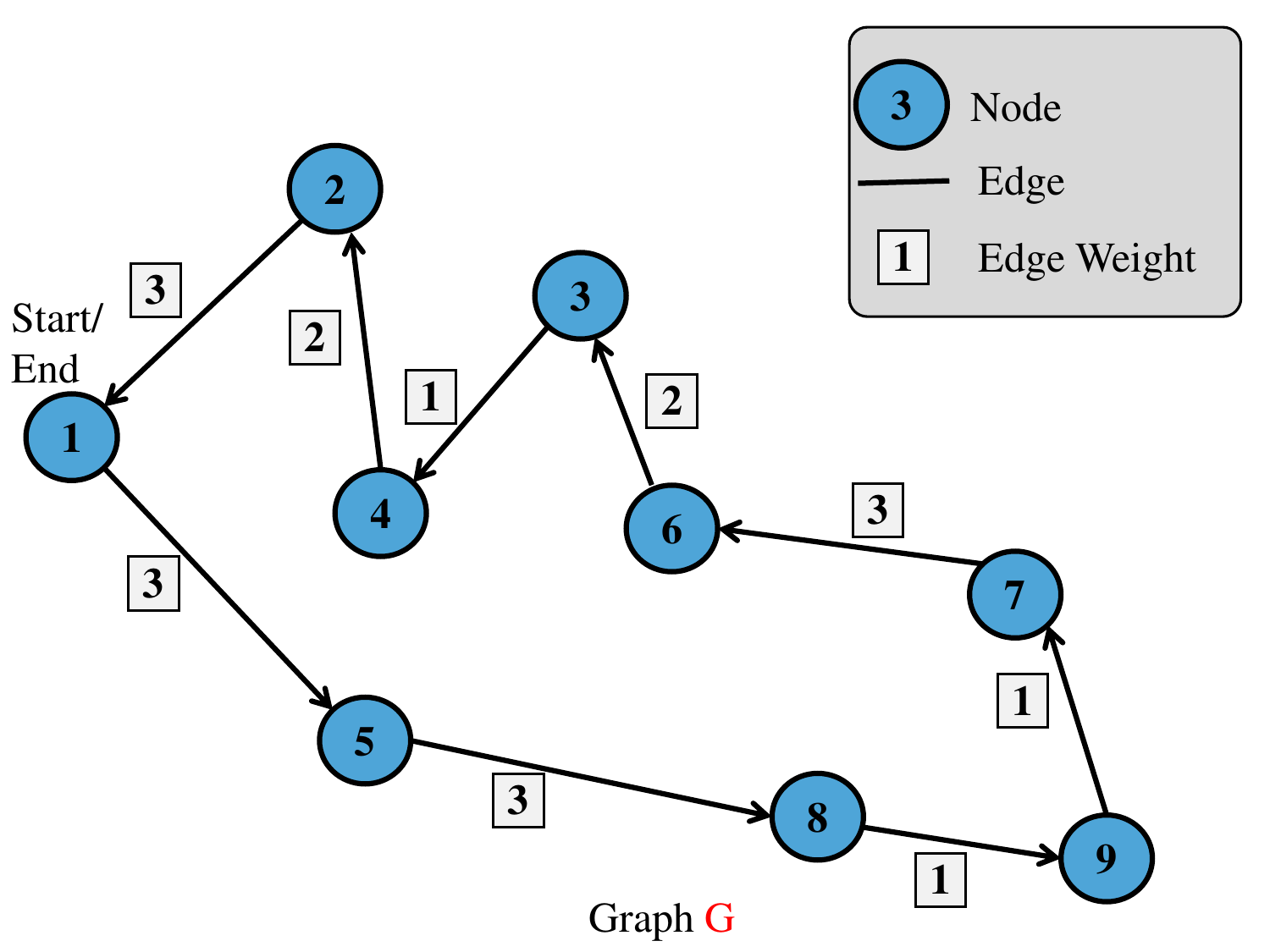}
		\caption{Illustration of a TSP Route\slash Tour Across a Graph.}
		\label{fig:TSP}
	\end{figure}
	
	The Traveling Salesman Problem (TSP) is a NP-hard problem in computer science, which tries to find the shortest possible route \slash tour that visits every node in a graph only once and then returns to the starting point. In a plant disease monitoring \slash management A-CPS, TSP can be used to compute the most efficient path for a drone or automated vehicle to cover all listed locations within a farm as part of daily routine \slash treatment.

	By representing the farm as a graph, where nodes are locations of interest and edges representing the distances or travel times between them, TSP algorithms can be applied to find the minimum route that visits all locations. Solving TSP helps in minimizes its travel distance or time required bu the drone to cover all locations, leading to more efficient operations. This can result in reduced fuel consumption, quicker task completion, and optimal resource management.


   \section{Hotspot Detection}
   \label{sec:Hotspot_Detection}

\begin{figure}[htbp]
		\centering
		\includegraphics[width=0.5\textwidth]{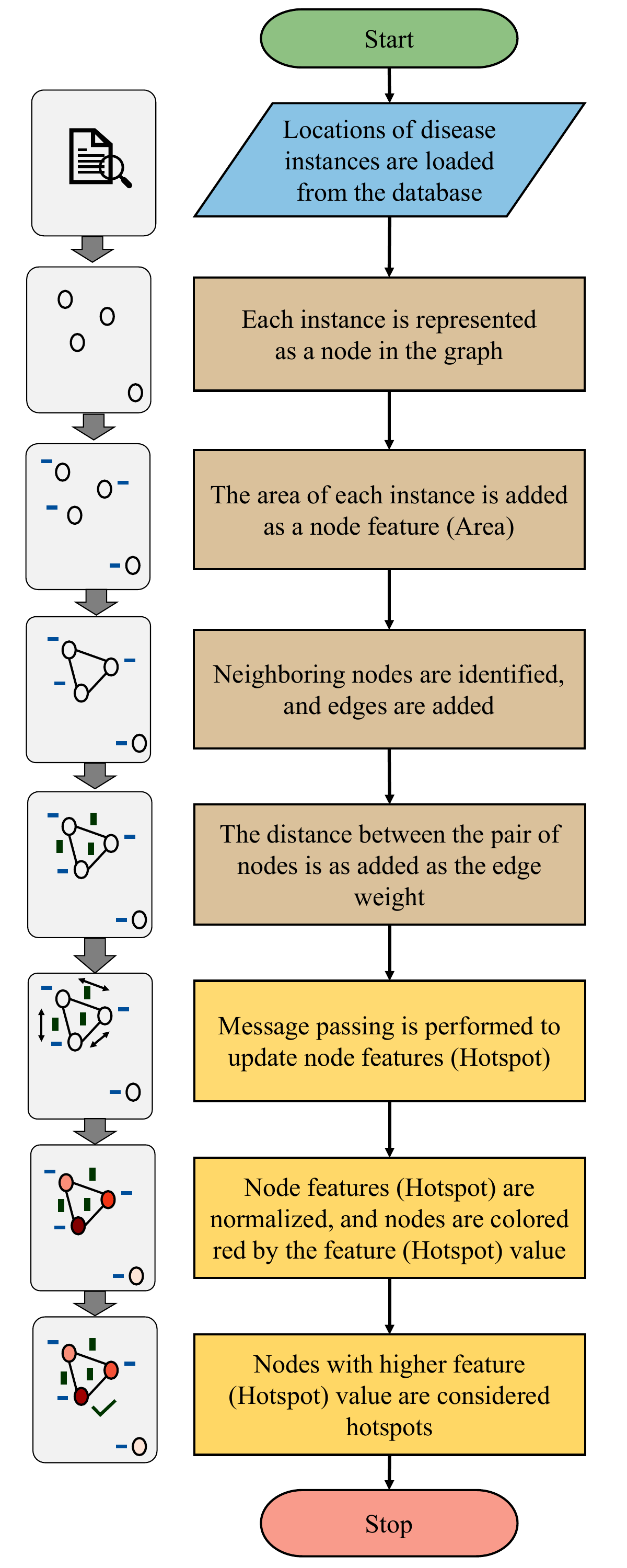}
		\caption{Hotspot Detection Flowchart.}
		\label{fig:Hotspot_Detection_Flowchart}
	\end{figure}

	The proposed "SprayCraft" system is focused on determining the optimal route for variable rate spraying and does not include disease detection within the farmland. It takes the coordinates of diseased locations in the farmland as input and uses them to achieve its routing and spraying objectives, as described in the following subsections. The flowchart in Fig \ref{fig:Hotspot_Detection_Flowchart} illustrates the method.
	
	\subsection{Graph Representation}
	\label{sec:Graph_Representation}		
	
	 In the proposed method for spatial analysis of disease distribution, each location of infection or disease is represented as a node in the graph. The process begins by initializing a graph structure $G$, which will be used to model the spatial relationships between different disease-affected areas. For each diseased location \slash segment, a node $u$ is created and its area is recorded as a feature $f_1$ of the node, is represented by the vector $h_u$. This feature captures both the severity of the disease and the potential presence of multiple affected plants at a given location.
	
\begin{figure}[htbp]
		\centering
		\includegraphics[width=0.4\textwidth]{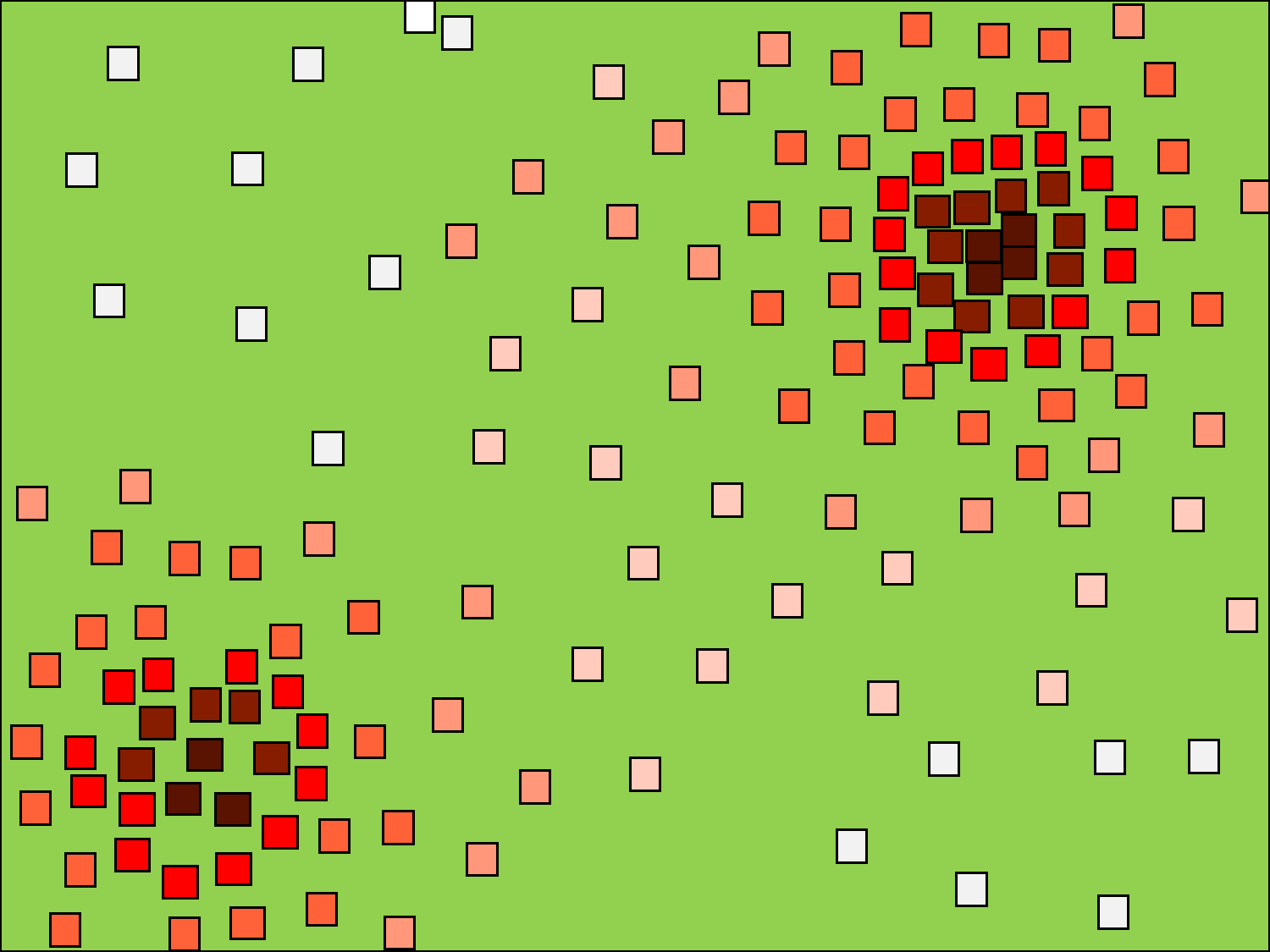}
		\caption{Presence of Multiple Disease Hotspots in Farmland.}
		\label{fig:Multiple_Disease_Propagation_Hotspots}
	\end{figure}

	Disease spread often occurs in dense plantations and can lead to multiple hotspots within a farmland as depicted by Fig \ref{fig:Multiple_Disease_Propagation_Hotspots}. If all nodes in the graph were considered neighbors, the entire farmland would be represented as a single cluster, which would fail to identify multiple hotspots. To address this, only nodes within a 25-meter radius are considered neighbors, accounting for pathogen interactions and wind dispersion. A weighted edge $(v_i, v_j,w)$ is created between neighboring nodes $i^{th}$ and $j^{th}$ nodes, with the weight being the distance between them.  is added between the pair. Consequently, all the diseased locations in the farmland are represented as a graph $G$ with nodes $V = \{v_1,v_2,v_3, \ldots, v_n\}$, edges $E = \{(v_i, v_j,w), (v_j, v_k,w'), \ldots\}$, features F = $\{f_1\}$ similar to the one in Fig \ref{fig:Graph}.
	
	\subsection{Hotspot Computation}
	\label{sec:Hotspot_Computation}
	
	Disease hotspot refers to a location with higher incidence when compared to its surroundings and act as sources of diseases. Accordingly, number of diseased plants closer to a hotspot is also higher. Hence, it can be inferred that a hotspot can be identified as a diseased location at the center of a high-intensity cluster. Given that every diseased location is represented as a node and edge is present between locations that are close to each other, To find a hotspot, at least $2$ hop neighbors of each location are to be analyzed. To accurately estimate the hotspots, in addition to considering the neighbors, the area of each neighbor location must be considered. So, we compute the sum of areas of the current node and all its neighbors to replace its current feature value with a newly computed value. As a result, the updated feature of every node will represent the combined area of the node and its immediate neighbors. Repeating the same action will update the node feature with the area of $2$ hop neighbors for the node and the feature value can be considered as a relative measure for the possibility of a node being a hotspot.

	In cases where nodes are densely clustered, the method described above will result in identical feature values for multiple nodes within the cluster. Since the probability of infection is higher for plants that are closer to the diseased plant, the distance between neighbors have to be considered to mitigate the risk of smothering of feature values.
	
	Now that the solution is defined, we will explore methods for its implementation. To update a node's current feature, we first identify all its neighbors. Since the distance between plants is inversely related to the probability of infection, we multiply each neighbor's feature value by the inverse of the corresponding edge weight. We then add the sum of these products to the current feature value of the node in Eqn. \ref{eq:Hotspot_detection}: 	
	
	\begin{equation}
		\label{eq:Hotspot_detection}
		h_u^{(k+1)} = \sum \left( h_u^{(k)}, \sum \left( \{ h_v^{(k)} * 1 / W(uv)  : v \in N(u) \} \right) \right) .
	\end{equation}	
	
	In Equation \ref{eq:Message_passing} the AGG function is a summation function ($\sum$), and the UPDATE function computes the summation ($\sum$) of the products of the feature value $h_v^{(k)}$ and the inverse of edge weight $W(uv)$ or neighbors $N(u)$ , which is represented in Equation \ref{eq:Hotspot_detection}. This indicates that applying message passing twice on the generated graph $G$ aids in identifying hotspots. To determine the neighbors of a node in the graph $G$, the Adjacency Matrix $A$ is constructed. An adjacency matrix is a $m \times m$ sized square matrix, where the number of nodes in the graph is $m$. An entry $A[i][j]$ is weight of the edge if an edge exists between nodes $i$ and if $j$ and 0 if no edge is present. The row $A[i]$ represents all edges that connect nodes  $i$ and if $j$ is a neighbor of $i$, $A[i][j]$ denotes weight $w$ of th edge $(v_i, v_j,w)$. Each node in graph $G$ has a single feature $f_1$ and the feature representation $h_u$ is its feature value. The matrix $H[i]$ representing the feature vectors of all nodes, is an $m \times 1$ matrix where $H[i]$ is the value of feature $f_1$ for $i^{th}$ node. Performing the matrix multiplication $1/A \times H$ produces a column vector where each entry $i^{th}$ row is the sum of the products of feature values and corresponding edge weights for all neighbors of the $i^{th}$ node, which is the aggregation function in Equation \ref{eq:Message_passing}. Therefore, Equation \ref{eq:Message_passing} can be expressed as Equation. \ref{eq:Hotspot_detection}:

	\begin{equation}
		\label{eq:Message_passing_matrix}
		H^{(k+1)} =  H^{(k)} + \left( A^{\circ -1} \times H^{(k)} \right).
	\end{equation}
	
	Applying this process twice to the generated graph will update the feature values as intended. The node features are normalized, the node feature can be considered as probability of the node being a hotspot and nodes with higher feature values are identified as hotspots. Additionally, the nodes are colored red based on their feature values to visually represent the hotspotness of each node.
	
	\section{Route Computation}
	\label{sec:Route_Computation}
	
	The path for an agricultural drone to effectively treat diseases is computed in two stages. The first stage involves finding the optimal tour path, which visits every node only once and returns to the starting point. Then, in the second stage, a Boustrophedon path is computed based on the previously calculated probability of hotspotness, to deliver a relative dosage of pesticide for each diseased instance. This two-stage approach ensures effective disease treatment while minimizing travel distance. The following subsections explain the methods in detail.

	\subsection{Tour Computation}
	\label{sec:Tour_Computation}
	
	For Traveling Salesman Problem (TSP), the time required to find an optimal solution grows exponentially with the number of nodes making it a NP-Hard problem. So, for large graphs finding an optimal solution becomes infeasible. Among the many algorithms attempting to solve this problem \cite{Dolias2022}, we use the Christofides Algorithm, which guarantees a solution that is at most 1.5 times the optimal TSP solution, balancing accuracy and computational efficiency.
	
	 First, Christofides' Algorithm, as detailed in Algorithm \ref{alg:Christofides_Algorithm}, finds a minimum spanning tree (MST) connecting nodes in the graph generated in Section \ref{sec:Graph_Representation} while keeping total edge weight minimum. Next, the algorithm identifies nodes with odd number of neighbors (odd degree nodes) in the MST and uses a minimum-weight perfect matching to pair these nodes. A perfect matching is a set of edges that connects all odd-degree nodes such that the total edge weight is minimized. Then, the algorithm combines the MST and the perfect matching to form a graph where all nodes have even degrees, resulting in a path that visits every edge only once and then goes to the starting point known Eulerian graph. Finally, the Eulerian Circuit is converted into a Hamiltonian Circuit, a path that visits each node exactly once, by shortcutting repeated nodes. Thus, the algorithm computes a near-optimal path for the drone that ensuring that every diseased location in the farmland is visited. 
	
	\begin{algorithm}[htbp]
		\small 
		\caption{Christofides Algorithm for TSP}
		\label{alg:Christofides_Algorithm}
		\KwIn{Graph $G = (V, E)$ with metric weights}
		\KwOut{Hamiltonian circuit $C$ approximating the TSP tour}
		\BlankLine
		Minimum spanning tree (MST) $T$ of $G$ is computed\;
		The set $O$ of vertices in $T$ with a degree that is odd is computed\;
		A minimum weight perfect matching $M$ in the subgraph induced by the set $O$ is computed\;
		The edges of $T$ and $M$ to form a multigraph $H$ are combined\;
		Eulerian circuit $E$ in $H$ is computed\;
		$E$ is converted to a Hamiltonian circuit $C$ by short-cutting repeated vertices\;
		\Return $C$
	\end{algorithm}

	\subsection{Boustrophedon path Computation}
	\label{sec:Boustrophedon_path_Computation}

\begin{figure}[htbp]
		\centering
		\includegraphics[width=0.4\textwidth]{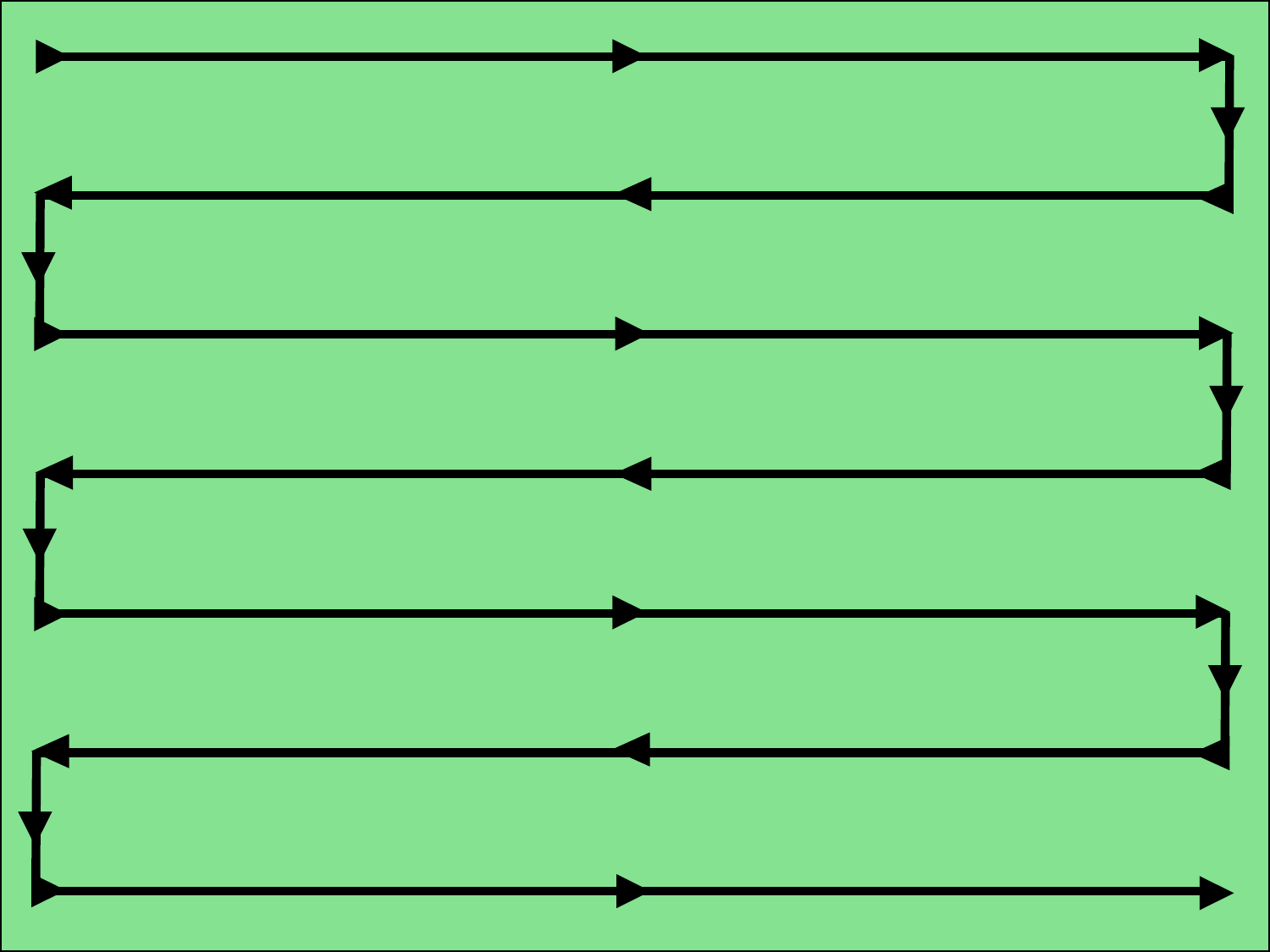}
		\caption{Boustrophedon path.}
		\label{fig:Boustrophedon_path}
	\end{figure}

	The start point of the drone is designated as the first coordinate in the flight path and each node in the graph is traversed as per the tour computed in Section \ref{sec:Tour_Computation}. While in each diseased location, to ensure effective coverage and thorough application of pesticide, the drone has to follow a suitable path allowing uniform coverage and spraying. Boustrophedon Path or Serpentine pattern involves traveling in parallel lines which alternate in direction after each pass as shown in Fig \ref{fig:Boustrophedon_path}. Since, this enables uniform pesticide spraying by a drone moving along the lines, we propose use of Boustrophedon Path.
	
\begin{figure}[htbp]
		\centering
		\includegraphics[width=0.5\textwidth]{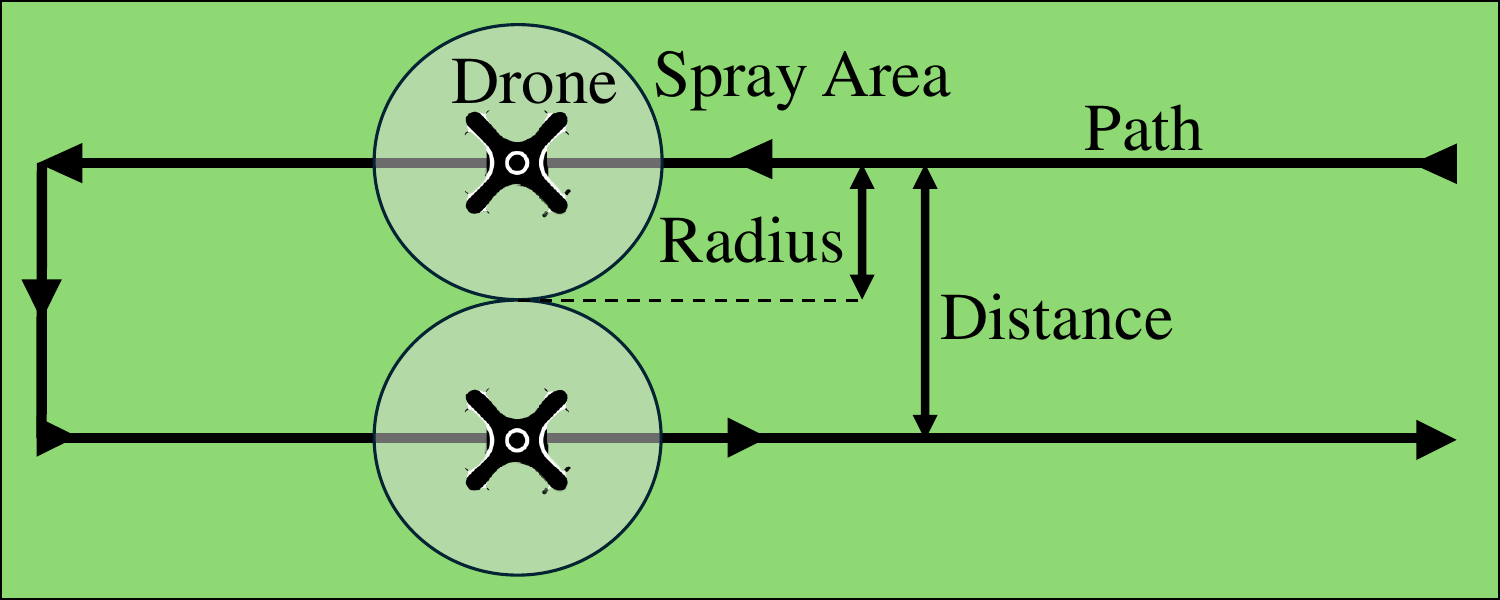}
		\caption{Relation Between Spray Radius and Path Spacing.}
		\label{fig:Full_Coverage}
	\end{figure}

	Considering image processing is used to detect presence of disease in the farmland which draws rectangular boxes around the identified objects, From the given coordinates of diseased locations, coordinates of each location are obtained to determine the four corners, height and width of the location. Since the drone sprays evenly on both sides of its path, the distance between two parallel paths is set to twice the spray radius of the drone as depicted in Fig \ref{fig:Full_Coverage}. Using the height and width of each diseased location, along with the specified distance between parallel paths, the Boustrophedon Path is computed as per the Algorithm \ref{alg:Boustrophedon_Path} and added to flight path. This ensures each area is fully covered, with paths spaced to match the drone's spraying radius, so there are no gaps or overlaps.

	An example usage of the Algorithm \ref{alg:Boustrophedon_Path} with $x_{\min} = 0$, $y_{\min} = 0$, $width = 5$, $height = 10$, and $distance = 2$ yields the following path points:
	\begin{itemize}
		\item $(0, 1)$, $(1, 1)$, $(2, 1)$, $(3, 1)$, $(4, 1)$, $(5, 1)$
		\item $(5, 3)$, $(4, 3)$, $(3, 3)$, $(2, 3)$, $(1, 3)$, $(0, 3)$
		\item $(0, 5)$, $(1, 5)$, $(2, 5)$, $(3, 5)$, $(4, 5)$, $(5, 5)$
		\item $(5, 7)$, $(4, 7)$, $(3, 7)$, $(2, 7)$, $(1, 7)$, $(0, 7)$
		\item $(0, 9)$, $(1, 9)$, $(2, 9)$, $(3, 9)$, $(4, 9)$, $(5, 9)$
		
	\end{itemize}

	After computing the paths for each location \slash instance, the starting point's coordinates are designated as the last coordinate in the flight path to return to the starting point. Thus, a complete tour is computed, beginning at the start point, traveling through each location, and returning to the start point.
	
	\begin{algorithm}[htbp]
		\caption{Boustrophedon Path Generation}
		\label{alg:Boustrophedon_Path}
		\small 
		\SetAlgoLined
		\DontPrintSemicolon
		\KwIn{$x\_min$, $y\_min$, $width$, $height$, $distance$}
		\KwOut{$path$}
		$path \leftarrow []$\;
		$buffer \leftarrow ((height \% distance) / 2) $\;
		$y\_start \leftarrow y\_min + ((distance/2) $ \textbf{if} $(height \% distance) == 0$ \textbf{else} $buffer)$\;
		$y\_end \leftarrow y\_min + height$\;
		$x\_end \leftarrow x\_min + width$\;
		$x \leftarrow x\_min$\;
		$y \leftarrow y\_start$\;
		\BlankLine
		\eIf{$height > distance$}{
			\While{$y \leq y\_end$}{
				\eIf{$x == x\_min$}{
					\tcp{Move to the right}
					\While{$x \leq x\_end$}{
						$path.append((x, y)$\;
						$x \leftarrow x + 1$\;
					}
					$x \leftarrow x - 1$\;
				}{
					\tcp{Move to the left}
					\While{$x \geq x\_min$}{
						$path.append((x, y))$\;
						$x \leftarrow x - 1$\;
					}
					$x \leftarrow x + 1$\;
				}
				$y \leftarrow y + full\_coverage$\;
			}
		}{
			\tcp{Height $\leq$ distance}
			\While{$x \leq x\_end$}{
				$y \leftarrow round(y\_min + (height / 2), 1)$\;
				$path.append((x, y))$\;
				$x \leftarrow x + 1$\;
			}
		}
		\Return{$path$}\;

	\end{algorithm}
	
\begin{figure*}[!htbp]
		\centering
		\includegraphics[width=0.98\textwidth]{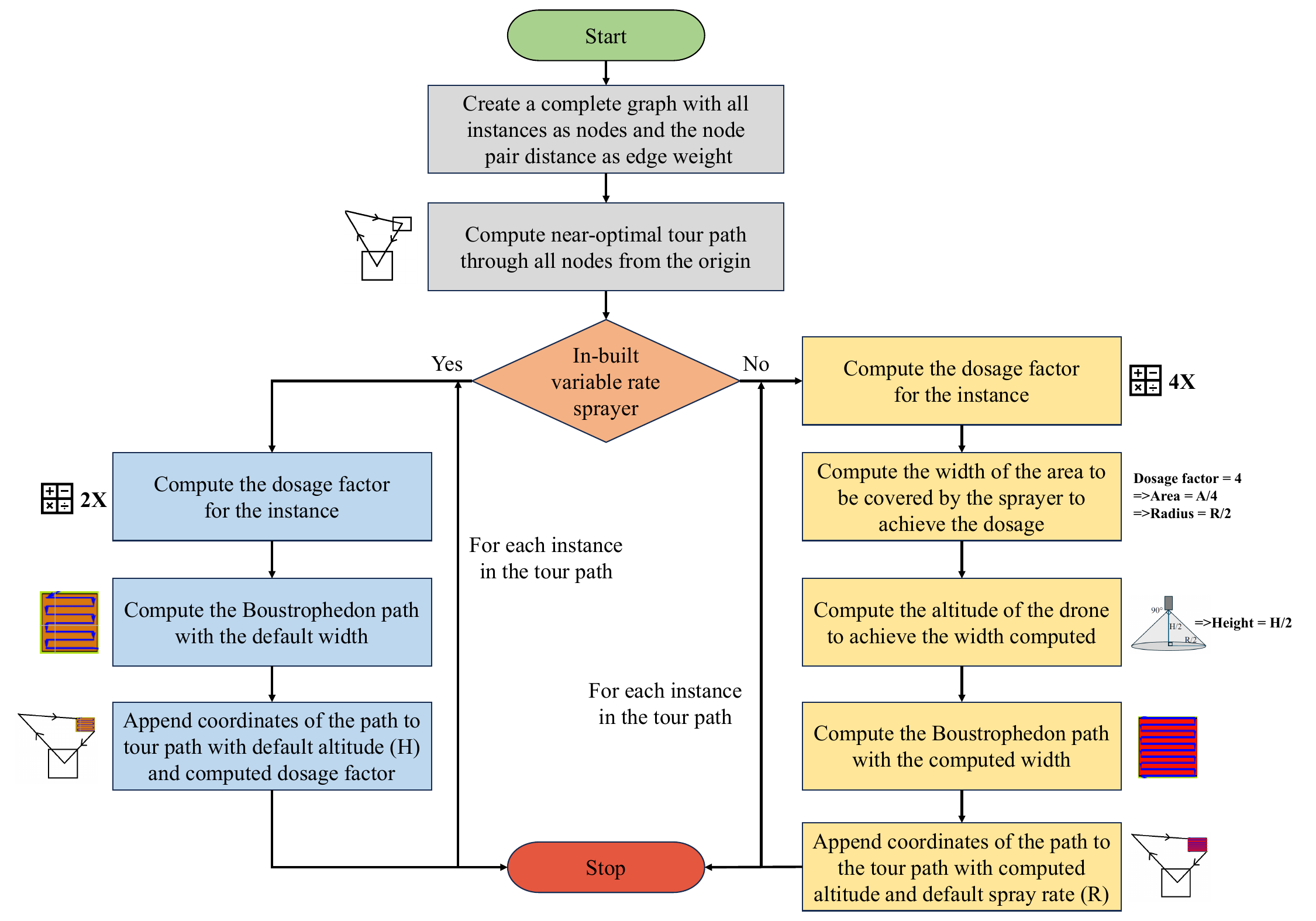}
		\caption{Path computation for Variable Rate Spraying.}
		\label{fig:Path_Computation_Flowchart}
	\end{figure*}

	The above path computation does not take into account the degree of infection or the probability of a location being a disease hotspot, and it cannot perform variable rate spraying. Therefore, we use the normalized feature values computed in Section \ref{sec:Hotspot_Computation}, which represent the degree of infection or probability of being a hotspot, to enable variable rate spraying. There are two types of spraying systems used in agriculture: one with a constant flow rate and one with a variable flow rate. Considering these setups, we propose an adaptable route computation method that takes into account the spray radius, the type of spray system, and the desired intensity factor for hotspots. This method computes the route as illustrated in Fig \ref{fig:Path_Computation_Flowchart} and following Sub Sections \ref{sec:With_Variable_Rate_Sprayer} and \ref{sec:Without_Variable_Rate_Sprayer}.
	
	\subsubsection{With Variable Rate Sprayer}
	\label{sec:With_Variable_Rate_Sprayer}
	
	In drones equipped with spray systems that have PID and PWM controls \cite{app8122482}, the flow rate can be adjusted according to the prescription map or the amount of pesticide needed while the drone maintains a constant altitude. For these systems, we input the spray width of the sprayer, flight height of the drone above the crop, the height of the crops, and the intensity factor for pesticide application in areas identified as primary hotspots (nodes with the highest feature values). For each location, the Boustrophedon path is computed with dimensions of location, twice the spray with as distance using Algorithm \ref{alg:Boustrophedon_Path}. The prescribed pesticide dosage is computed by multiplying the normalized feature value by the base dosage. The flight altitude is calculated by adding the crop height to the flight height above the crop. These flight altitude and prescribed dosage values are then integrated into the previously computed path coordinates. As a result, the drone path is represented as a 4-point coordinate array: the first three points represent the X, Y, and Z coordinates of the flight path, while the fourth point indicates the factor by which the base flow rate of the spray system should be adjusted based on the hotspot probability of the location to achieve variable rate spraying.

	\subsubsection{Without Variable Rate Sprayer}
	\label{sec:Without_Variable_Rate_Sprayer}
	
\begin{figure}[!htbp]
		\centering
		\includegraphics[width=0.6\textwidth]{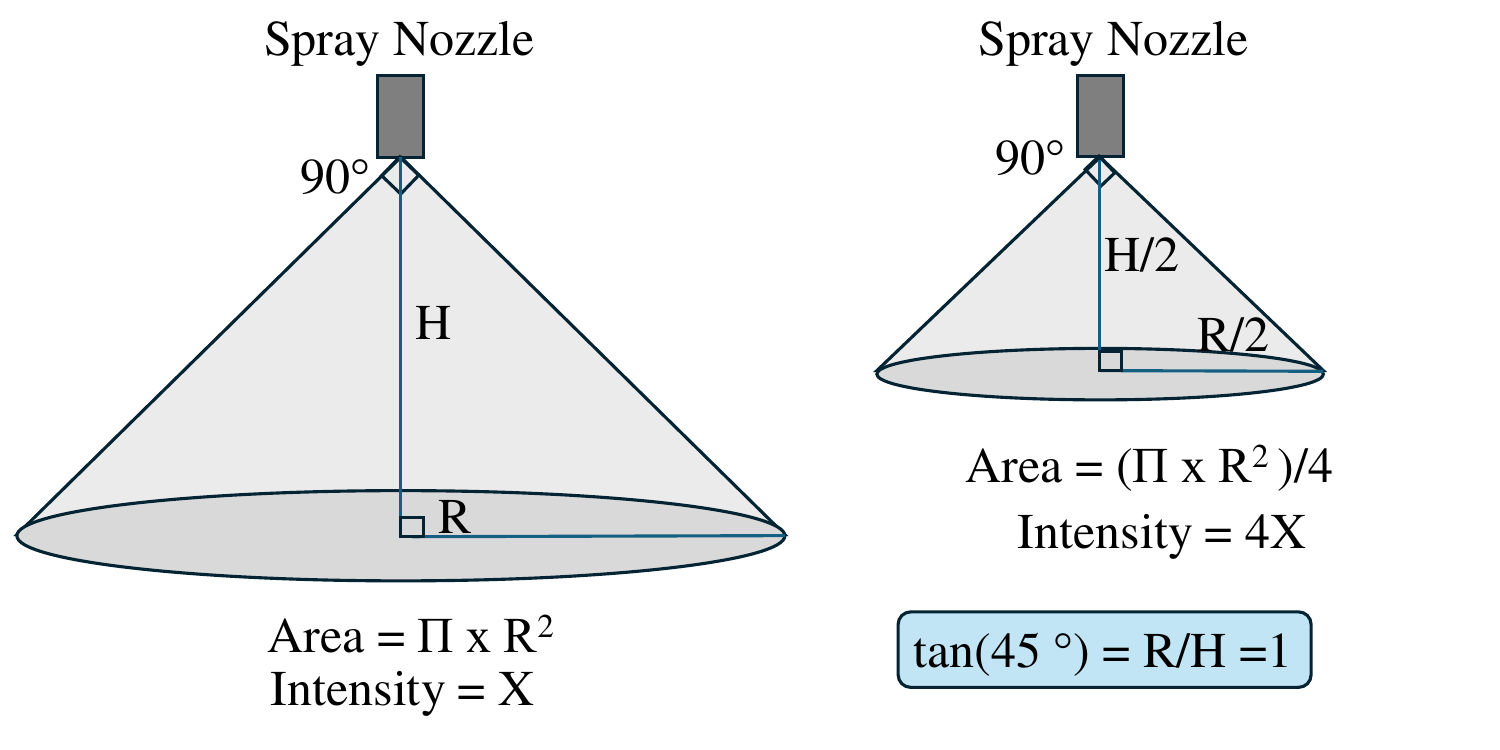}
		\caption{Relation Between Spray Radius and Spray Height.}
		\label{fig:Nozzle_Height_Area}
	\end{figure}

	For systems that do not have built-in variable rate mechanisms, we propose adjusting the flight height to control pesticide application. In spray systems, the nozzle disperses the liquid in a conical pattern, with the spray angle fixed. By altering the flight height, the effective coverage area of the drone decreases \cite{drones6120383}. Consequently, with a constant flow rate, the reduction in spray area leads to an increase in the amount of pesticide deposited, as illustrated in Fig \ref{fig:Nozzle_Height_Area}. This method allows for variable pesticide concentration in targeted areas despite the lack of variable flow rate control.
	
	 For these systems, we input the height of the crops, the intensity factor for pesticide application in areas identified as primary hotspots, and the flight height of the drone above the crop to ensure the standard dosage ($1$x) of pesticide is applied. For each instance\slash location, the intensity factor for pesticide application is computed similarly to the method used for systems with variable rate sprayers. In addition, the height of flight needed to achieve the computed intensity factor and the corresponding spray radius are determined. For example, consider a spray nozzle with a spray angle of 90\degree, as shown in Fig \ref{fig:Full_Coverage} .To achieve an intensity factor of 4, the coverage area must be reduced by a factor of 4, which means the spray radius needs to decrease by a factor of 2. Since  $\tan$(45\degree) is $1$, the height at which the sprayer system is positioned above the plant must also be reduced by a factor of 2. The Boustrophedon path is computed with dimensions of location, twice the computed spray with as distance using Algorithm \ref{alg:Boustrophedon_Path}.

	The flight altitude is calculated by adding the crop height to the computed flight height. This flight altitude, along with the constant flow rate, is then integrated into the previously computed path coordinates. As a result, the drone path is represented as a 4-point coordinate array: the first three points represent the X, Y, and Z coordinates of the flight path to achieve variable rate spraying based on the hotspot probability of the location.
	
	\subsection{With GPS coordinates}
	\label{sec:With_GPS_coordinates}
	
	To ensure compatibility with geo-sensing imagery that produces results in GPS coordinates and agricultural drones with GPS routing capabilities, we integrate methods to accept GPS coordinates and generate GPS coordinates for routing. To use the methods described in Sections \ref{sec:Tour_Computation} and \ref{sec:Boustrophedon_path_Computation}, we need to compute the dimensions of diseased locations in meters and then calculate GPS coordinates for routing based on these dimensions. The following formulas are used for adapting GPS coordinates.

	The distance \(d\) between two points with latitudes \(\phi_1\) and \(\phi_2\), and longitudes \(\lambda_1\) and \(\lambda_2\) is given by:

   \begin{equation}
   	\begin{aligned}
   		d = 2r \cdot \arcsin \Bigg( \Bigg[ & \sin^2 \left( \frac{\phi_2 - \phi_1}{2} \right) + \\
   		& \cos(\phi_1) \cdot \cos(\phi_2) \cdot \sin^2 \left( \frac{\lambda_2 - \lambda_1}{2} \right) \Bigg]^{1/2} \Bigg)
   	\end{aligned}
   \end{equation}

	A point that is \( x \) meters east of a given point \((\phi, \lambda)\) can be found using:
	
	\begin{equation}
	\lambda_{\text{new}} = \lambda + \frac{x}{r \cdot \cos(\phi)} \cdot \frac{180}{\pi}
    \end{equation}
	
	A point that is \( x \) meters west of a given point \((\phi, \lambda)\) can be found using:
	
	\begin{equation}
	\lambda_{\text{new}} = \lambda - \frac{x}{r \cdot \cos(\phi)} \cdot \frac{180}{\pi}
	\end{equation}
	
	A point that is \( x \) meters north of a given point \((\phi, \lambda)\) can be found using:
	
	\begin{equation}
	\phi_{\text{new}} = \phi + \frac{x}{r} \cdot \frac{180}{\pi}
	\end{equation}
	
	A point that is \( x \) meters south of a given point \((\phi, \lambda)\) can be found using:
	
	\begin{equation}
	\phi_{\text{new}} = \phi - \frac{x}{r} \cdot \frac{180}{\pi}
	\end{equation}

	where:
	
	\[
	r \text{ is Earth's radius (6,371,000 m)}
	\]

	\section{Experimental Verification}
	\label{sec:Experimental_Verification}

\begin{figure}[!htbp]
		\centering
		\begin{minipage}{0.49\textwidth}
			\centering
			\includegraphics[width=1\textwidth]{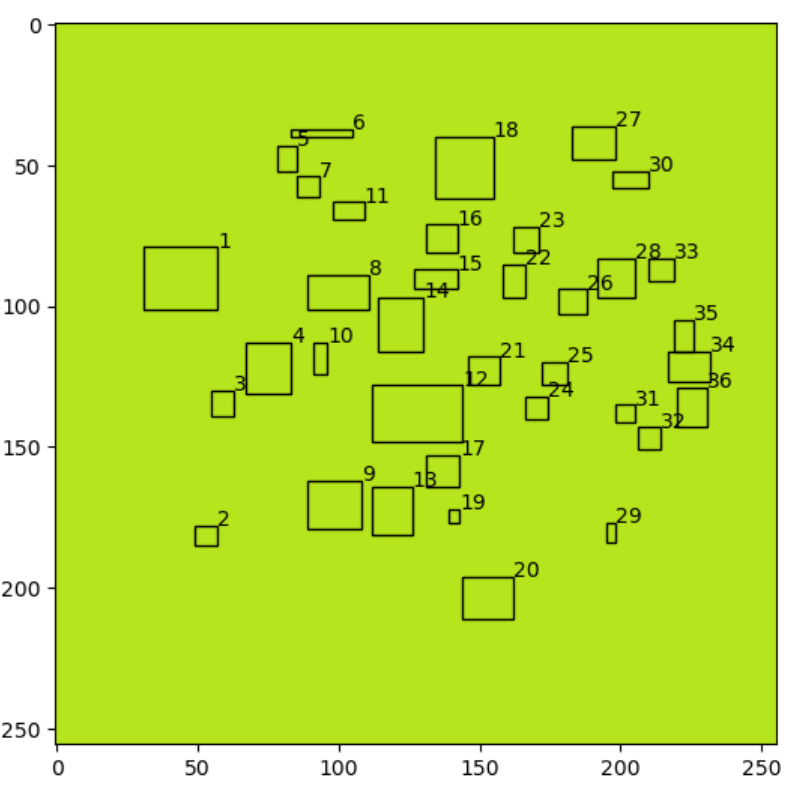}
			\caption{Diseased locations in the farmland.}
			\label{fig:Infected_Instances_1}
		\end{minipage}
		\hfill
		\begin{minipage}{0.49\textwidth}
			\centering
			\includegraphics[width=1\textwidth]{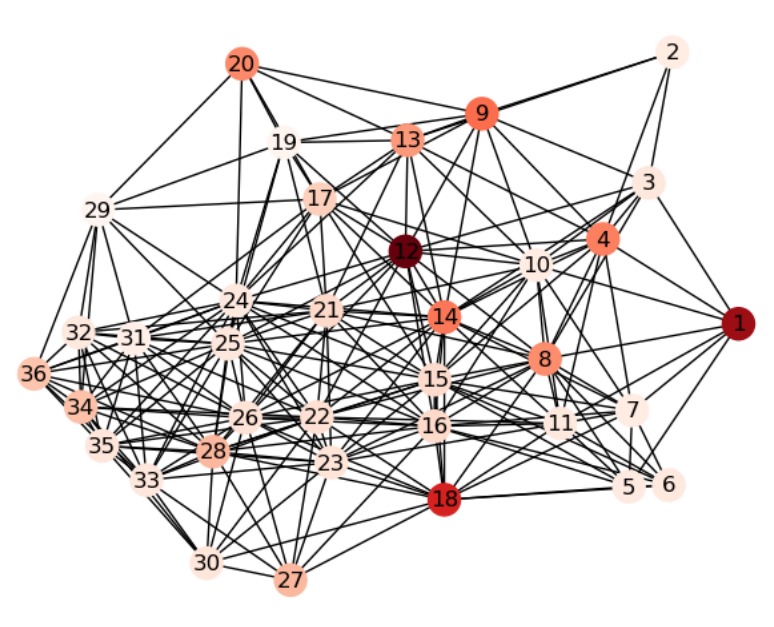}
			\caption{Graph representation of diseased locations.}
			\label{fig:Graph_Representation_1}
		\end{minipage}
	\end{figure}

	The graph-based solution described in Section \ref{sec:Hotspot_Detection}, \ref{sec:Route_Computation} was implemented using Python and the NetworkX library, which facilitates the creation and manipulation of graphs. For experimental verification, we generated synthetic data consisting coordinates of diseased locations of different sizes in a farmland. These locations were used to simulate disease sites identified by the disease detection mechanisms. In this section, we present the results for two sets of such data. Image of a farmland with diseased instances in it identified is shown in Fig \ref{fig:Infected_Instances_1}. Each instance in the provided data is represented as a node in the graph constructed according to the proposed method. The nodes are colored as per their node features and so the nodes of larger locations are in darker color in Fig \ref{fig:Graph_Representation_1}.

\begin{figure}[!htbp]
		\centering
		\includegraphics[width=0.7\textwidth]{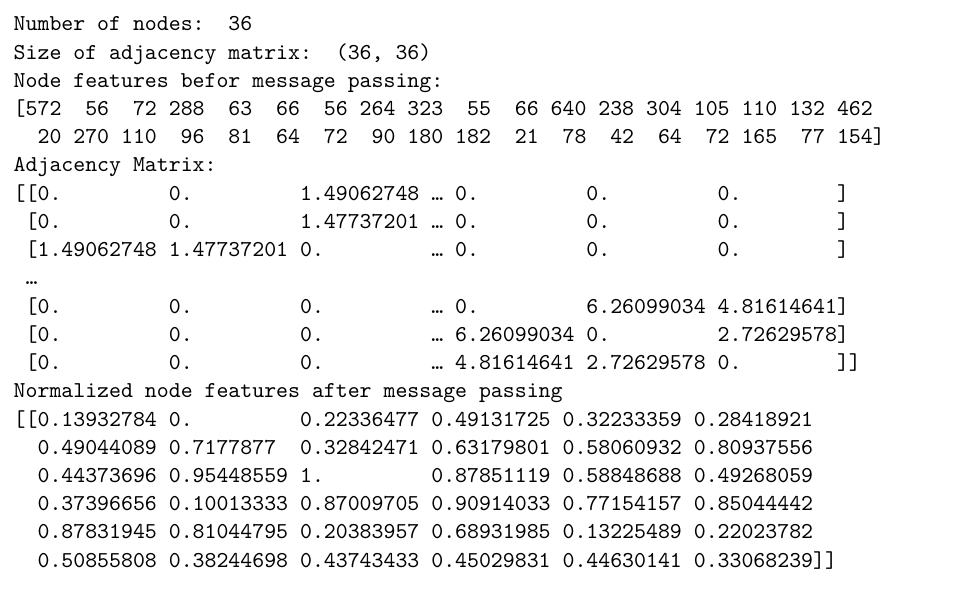}
		\caption{Node features before and after message passing.}
		\label{fig:Graph_Info_1}
	\end{figure}

	Message passing is then performed on the generated graph to learn about the neighbors and update their features, as shown in Fig \ref{fig:Graph_Info_1}.
	
\begin{figure}[!htbp]
		\centering
		\begin{minipage}{0.49\textwidth}
			\centering
			\includegraphics[width=1\textwidth]{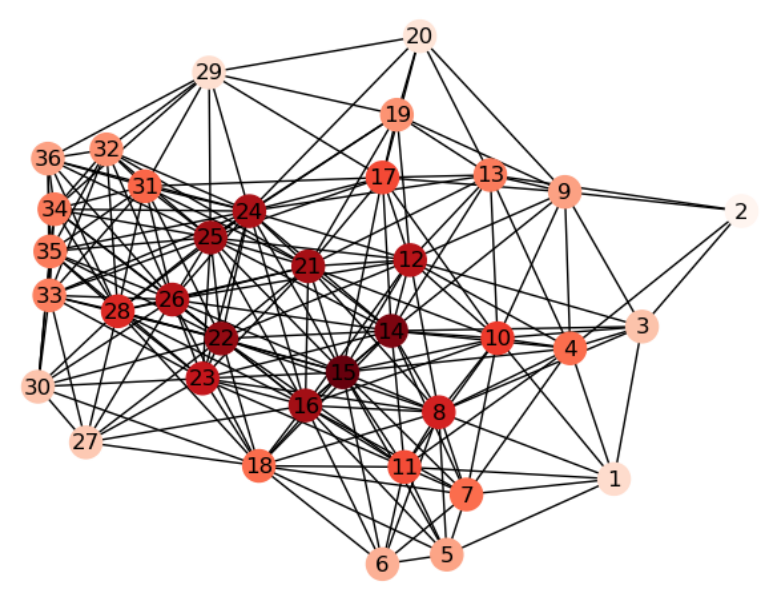}
			\caption{Graph representation after applying Message Passing algorithm.}
			\label{fig:Graph_Message_Passing_1}
		\end{minipage}
		\hfill
		\begin{minipage}{0.49\textwidth}
			\centering
			\includegraphics[width=1\textwidth]{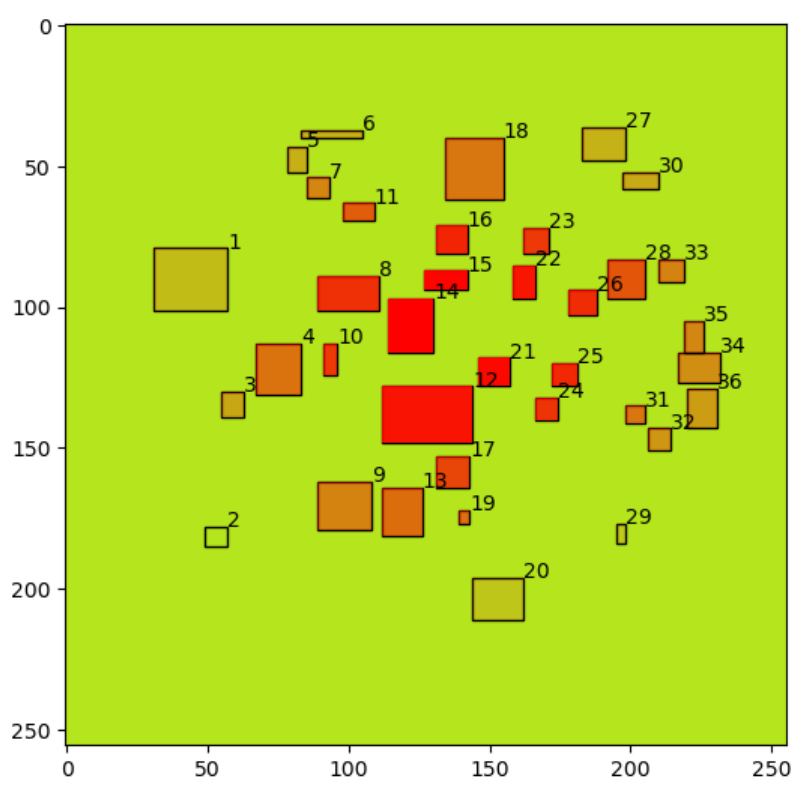}
			\caption{Representation of hotspots, colored by their probability.}
			\label{fig:Hotspots_Identified_1}
		\end{minipage}
	\end{figure}

	 As the intention of the proposed message passing is to identify disease hotspots, locations/nodes with more neighbors and larger neighboring nodes in their proximity attained higher feature values. Consequently, these nodes $15$,$14$ are depicted in darker red in Fig \ref{fig:Graph_Message_Passing_1} and Fig \ref{fig:Hotspots_Identified_1} .	
	
\begin{figure}[!htbp]
		\centering
		\includegraphics[width=0.8\textwidth]{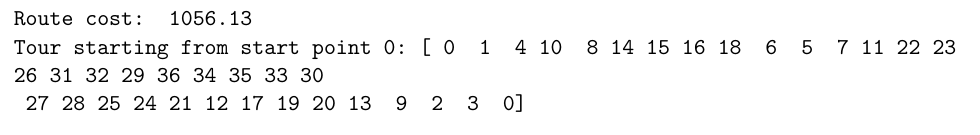}
		\caption{Computed Tour for the Graph generated.}
		\label{fig:Computed_Tour_1}
	\end{figure}

	From the graph generated, minimum spanning tree has been computed and near optimal tour path shown in Fig \ref{fig:Computed_Tour_1} is computed by Christofides approximation. This tour path starts from one corner of the farmland designated as starting point, visits every node once and returns to the same corner ending the tour.

\begin{figure}[!htbp]
		\centering
		\includegraphics[width=0.4\textwidth]{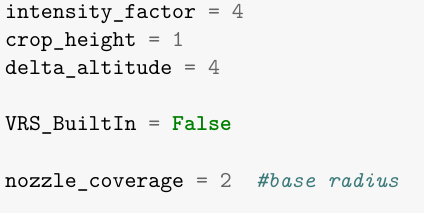}
		\caption{Parameters for drone without variable rate spraying.}
		\label{fig:Input_Params_1_1}
	\end{figure}

	To compute the Boustrophedon Path, we need to input values for the height of the crop, the flight height above the crop for base rate spraying, the radius of the area covered by the sprayer, and the intensity factor required at the primary hotspot in terms of the base rate. The type of spray system must also be specified using the "VRS\_BuiltIn" parameter. If "VRS\_BuiltIn" is set to "False", it means the system cannot perform variable rate spraying, and the method must compute the relative flight height to achieve variable rate spraying. If set to "True", the system can adjust the flow rate of the nozzle according to the prescription map. To generate path for a drone without variable rate sprayer, we set parameters as shown in Fig \ref{fig:Input_Params_1_1}.
	
\begin{figure}[!htbp]
		\centering
		\begin{minipage}{0.49\textwidth}
			\centering
			\includegraphics[width=1\textwidth]{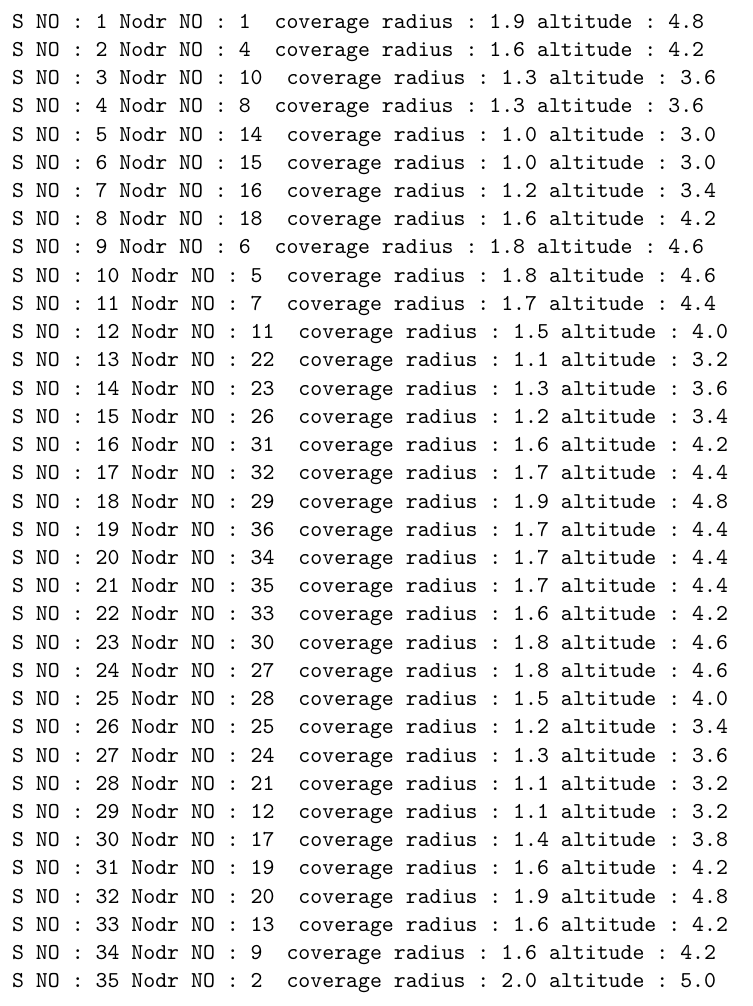}
			\caption{Path computation for constant rate sprayer.}
			\label{fig:Route_Intensity_1_1}
		\end{minipage}
		\hfill
		\begin{minipage}{0.49\textwidth}
			\centering
			\includegraphics[width=1\textwidth]{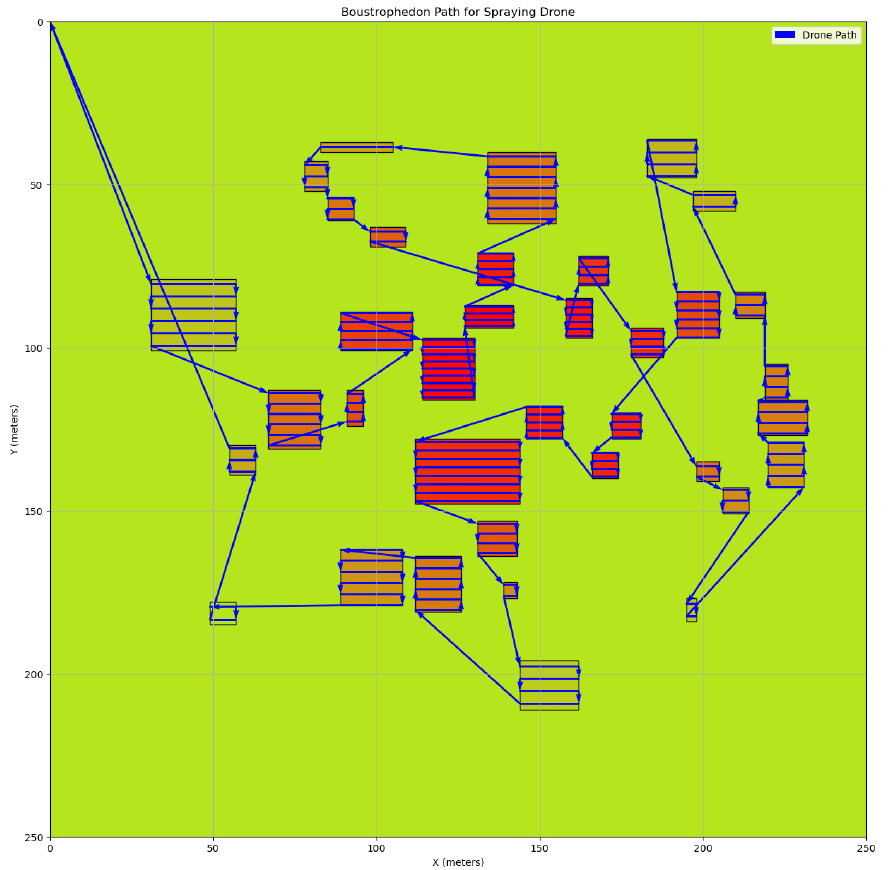}
			\caption{Path computed for Constant Rate Sprayer as per given parameters.}
			\label{fig:Drone_Path_Hotspot_1_1}
		\end{minipage}
	\end{figure}

	With the given parameters, the path is computed, traversing all locations as shown in Fig \ref{fig:Computed_Tour_1}. It calculates the flight altitude for the Boustrophedon path as depicted in Fig \ref{fig:Route_Intensity_1_1}, applying pesticide to the plants based on their probability of being a hotspot and the set intensity factor. The complete path computed for variable rate precision spraying, overlaid with the diseased locations is shown in Fig \ref{fig:Drone_Path_Hotspot_1_1}. The spacing between parallel paths in the diseased locations decreases with an increase in the probability of being a hotspot, as the drone adjusts its altitude to reduce the target area and deposit a greater amount of pesticide.

\begin{figure}[!htbp]
		\centering
		\begin{minipage}{0.49\textwidth}
			\centering
			\includegraphics[width=0.82\textwidth]{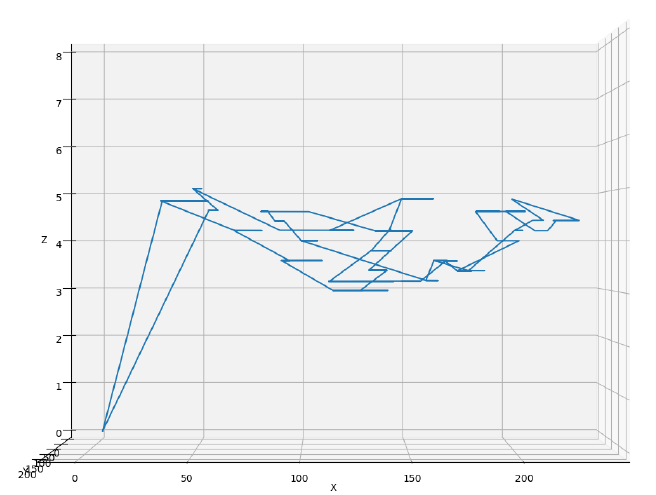}
			\caption{Map of Flight Altitude for the path computed.}
			\label{fig:Height_Map_1_1}
		\end{minipage}
		\hfill
		\begin{minipage}{0.49\textwidth}
			\centering
			\includegraphics[width=0.95\textwidth]{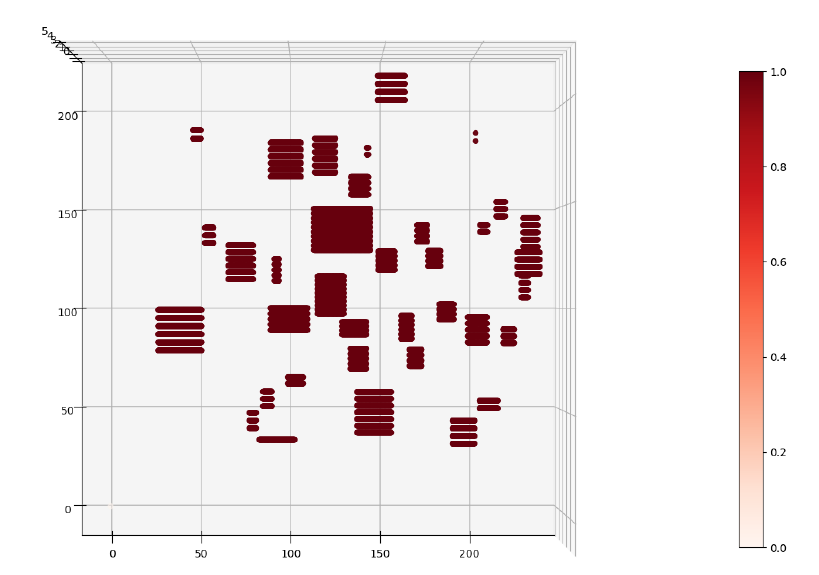}
			\caption{Heat map of sprayer flow rate for Constant Rate Sprayer.}
			\label{fig:Heat_Map_1_1}
		\end{minipage}
	\end{figure}

	This is also reflected in the flight altitude map shown in Fig \ref{fig:Height_Map_1_1}, where the drone flies lower over hotspot locations.
	
	Since the drone considered in this experiment does not have a variable rate sprayer, the flow rate remains constant throughout the flight, as shown in Fig \ref{fig:Heat_Map_1_1}.
	
\begin{figure}[!htbp]
		\centering
		\includegraphics[width=0.4\textwidth]{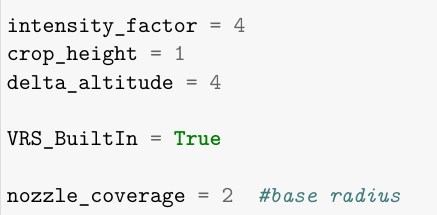}
		\caption{Parameters for drone with variable rate spraying.}
		\label{fig:Input_Params_1_2}
	\end{figure}

	
	Now, a drone with an in-built variable rate sprayer is considered. It uses the prescription map shown in Fig \ref{fig:Hotspots_Identified_1}, and the parameters are set according to Fig \ref{fig:Input_Params_1_2}.
	
\begin{figure}[!htbp]
		\centering
		\begin{minipage}{0.49\textwidth}
			\centering
			\includegraphics[width=1\textwidth]{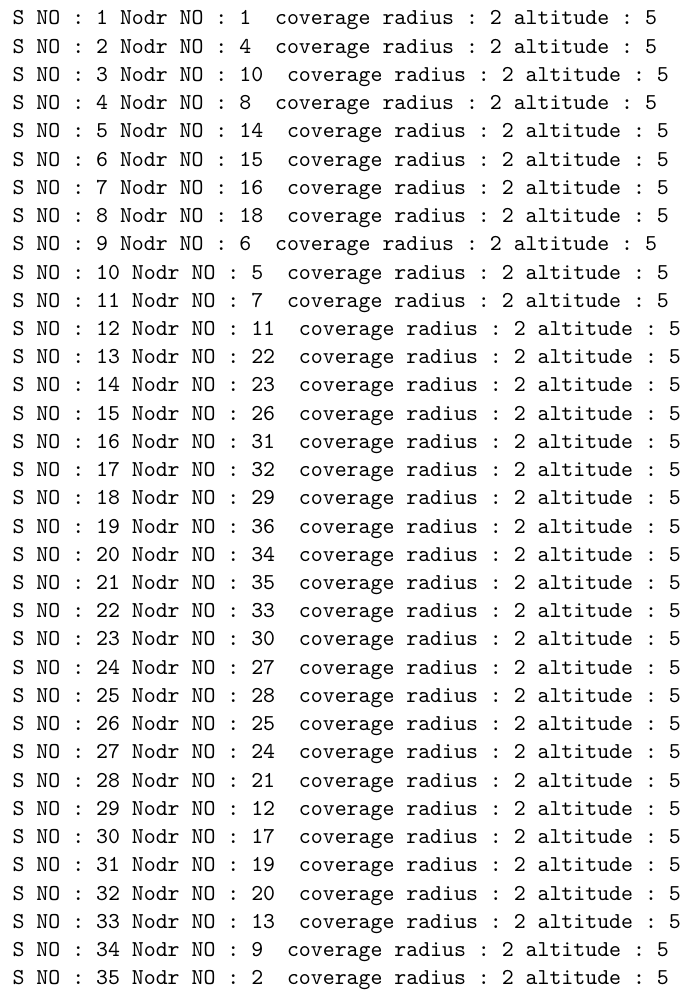}
			\caption{Path computation for variable rate sprayer.}
			\label{fig:Route_Intensity_1_2}
		\end{minipage}
		\hfill
		\begin{minipage}{0.49\textwidth}
				\centering
			\includegraphics[width=1\textwidth]{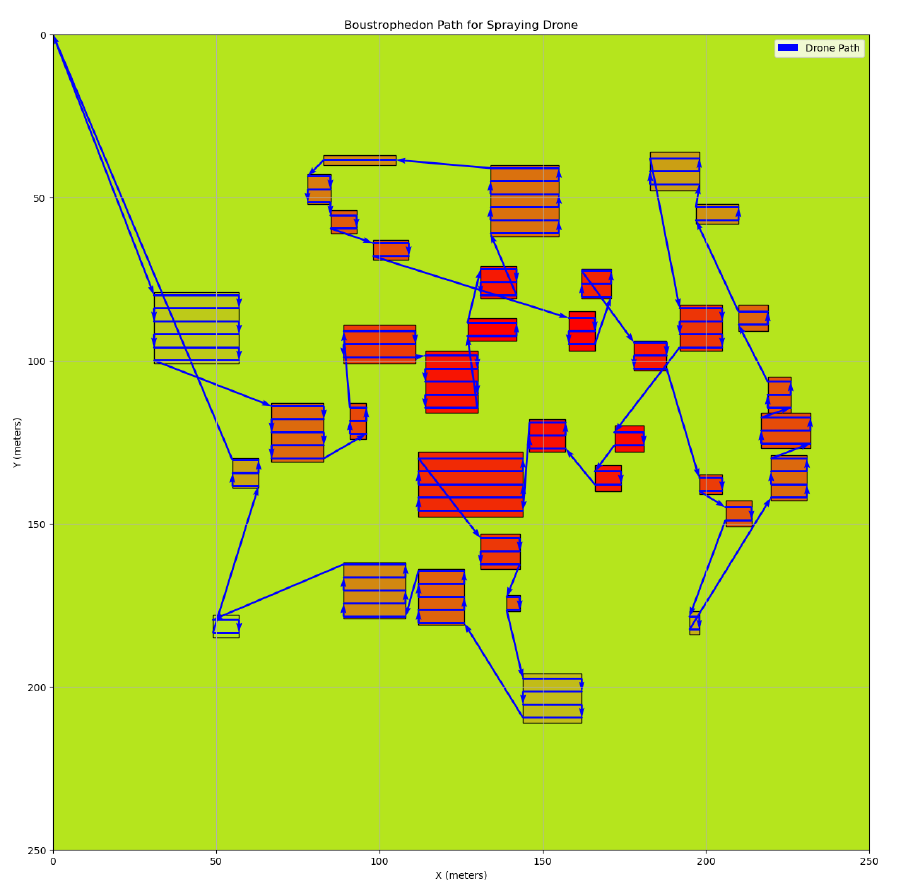}
			\caption{Path computed for Variable Rate Sprayer as per given parameters.}
			\label{fig:Drone_Path_Hotspot_1_2}
		\end{minipage}
	\end{figure}

\begin{figure}[!htbp]
		\centering
		\begin{minipage}{0.49\textwidth}
			\centering
			\includegraphics[width=0.82\textwidth]{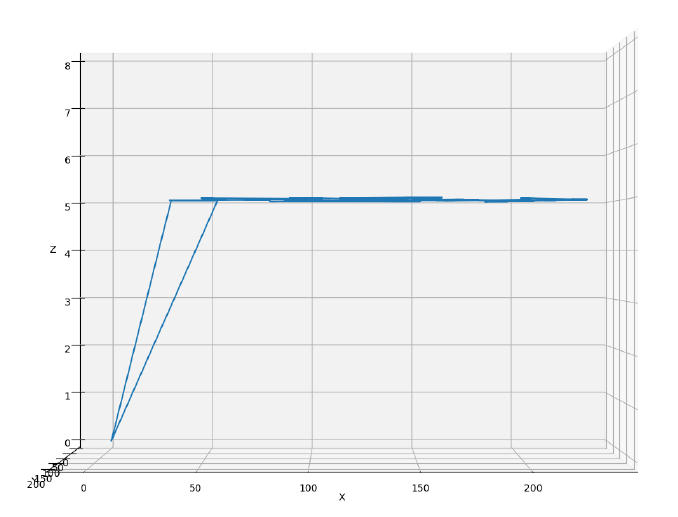}
			\caption{Map of Flight Altitude for the path computed.}
			\label{fig:Height_Map_1_2}
		\end{minipage}
		\hfill
		\begin{minipage}{0.49\textwidth}
			\centering
			\includegraphics[width=0.95\textwidth]{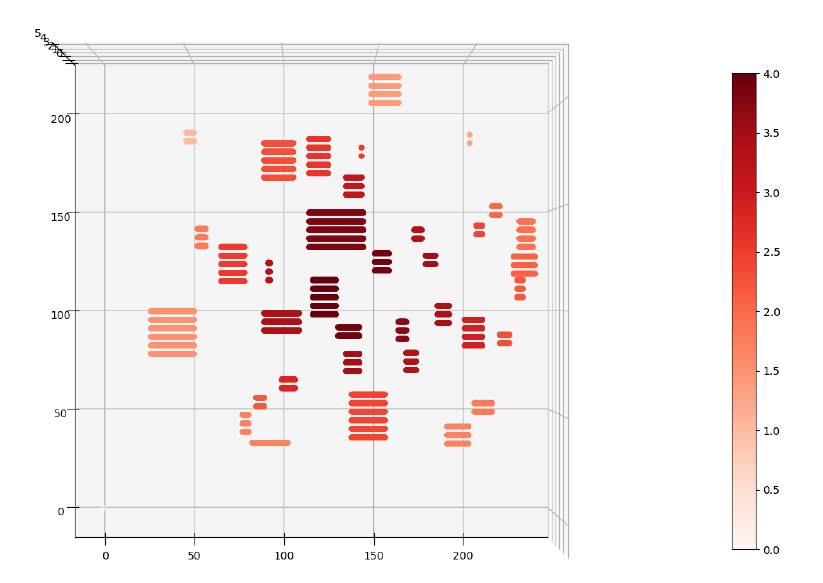}
			\caption{Heat map of sprayer flow rate for Variable Rate Sprayer.}
			\label{fig:Heat_Map_1_2}
		\end{minipage}
	\end{figure}

	With the given parameters, the path is computed, traversing all locations as shown in Fig \ref{fig:Computed_Tour_1}. And since it has ability to change the flow rate, altitude of flight remains constant as in Fig \ref{fig:Route_Intensity_1_2}. The complete path computed, overlaid with the diseased locations is shown in Fig \ref{fig:Drone_Path_Hotspot_1_1}. Since the system has an in-built variable rate mechanism, the spacing between parallel paths in Fig \ref{fig:Drone_Path_Hotspot_1_2} and the altitude of flight in Fig \ref{fig:Height_Map_1_2} remain constant.

\begin{figure}[!htbp]
		\centering
		\begin{minipage}{0.49\textwidth}
			\centering
			\includegraphics[width=0.8\textwidth]{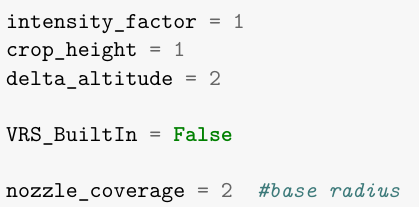}
			\caption{Parameters for drone to perform constant rate spraying.}
			\label{fig:Input_Params_1_3}
		\end{minipage}
		\hfill
		\begin{minipage}{0.49\textwidth}
			\centering
			\includegraphics[width=1\textwidth]{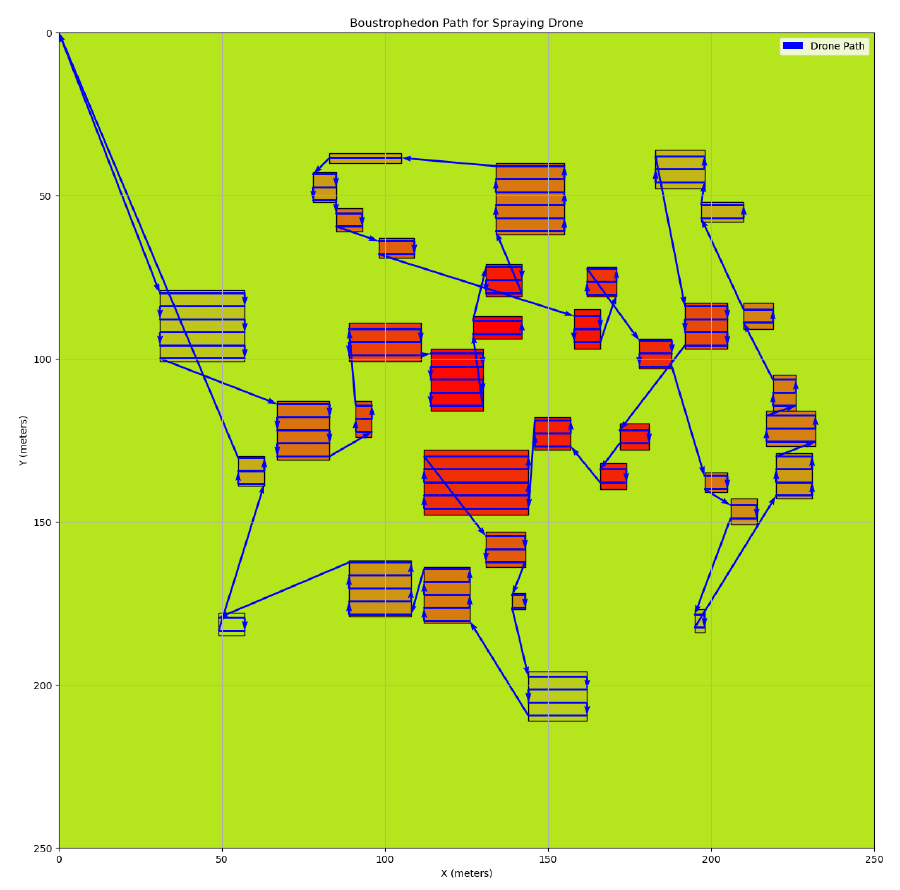}
			\caption{Path computed to perform constant rate spraying.}
			\label{fig:Drone_Path_Hotspot_1_3}
		\end{minipage}
	\end{figure}

	In some cases, farmers choose to spray the same amount of pesticide across the farm, and under such conditions, the flight altitude and flow rate must be maintained constant.  A case where the intensity factor is set to 1, meaning that the pesticide dosage is uniform throughout the farmland is illustrated in Fig \ref{fig:Input_Params_1_3}.
	
\begin{figure}[!htbp]
		\centering
		\begin{minipage}{0.49\textwidth}
			\centering
			\includegraphics[width=0.82\textwidth]{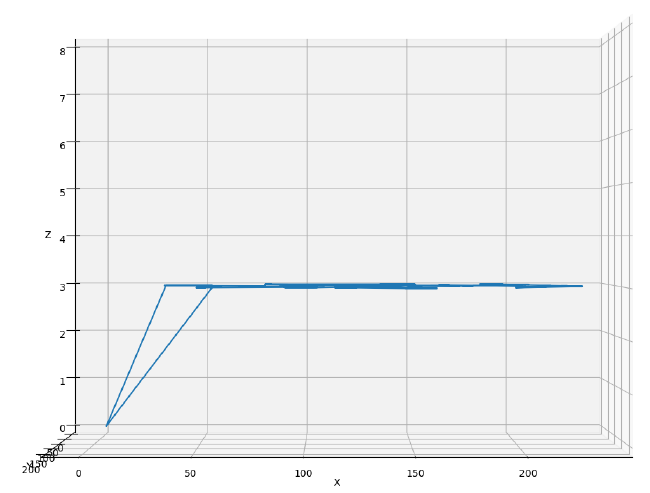}
			\caption{Map of Flight Altitude for the path computed.}
			\label{fig:Height_Map_1_3}
		\end{minipage}
		\hfill
		\begin{minipage}{0.49\textwidth}
			\centering
			\includegraphics[width=0.95\textwidth]{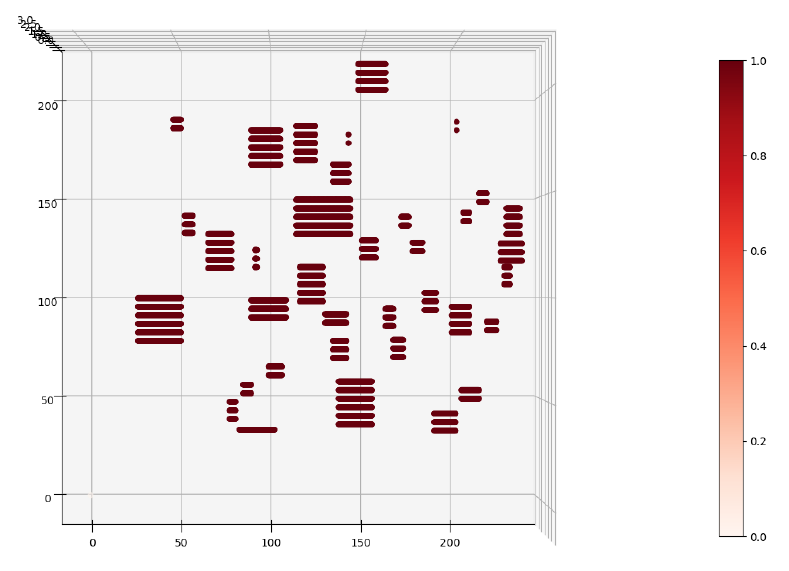}
			\caption{Heat map of sprayer flow rate for Variable Rate Spraying.}
			\label{fig:Heat_Map_1_3}
		\end{minipage}
	\end{figure}

	The path computed to achieve constant rate spraying is plotted in Fig \ref{fig:Drone_Path_Hotspot_1_3}, since its height and target area do not change with probability of being hotspotness, the spacing between the paths in Fig \ref{fig:Drone_Path_Hotspot_1_3} and the altitude of flight in Fig \ref{fig:Height_Map_1_3} remain constant.


	In the case of a sprayer with an in-built variable rate system, although the spacing between paths remains the same, the flow rate varies with the location. In contrast, for constant rate spraying, both the spacing between paths and the flow rate remain constant, as depicted in Figures \ref{fig:Drone_Path_Hotspot_1_3}, \ref{fig:Heat_Map_1_3}.
		
\begin{figure}[!htbp]
		\centering
		\begin{minipage}{0.49\textwidth}
			\centering
			\includegraphics[width=0.7\textwidth]{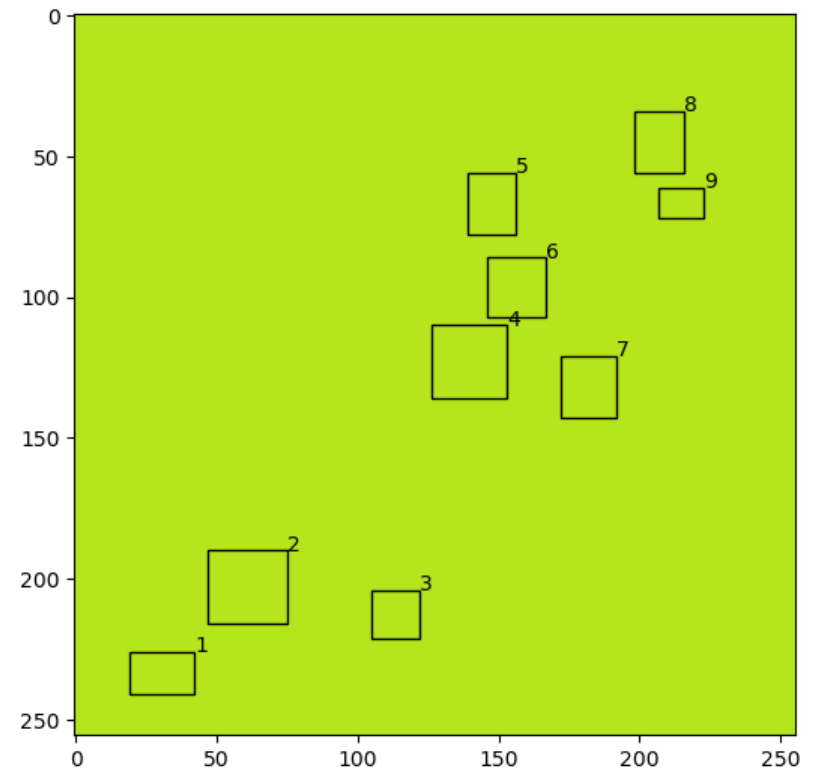}
			\caption{Diseased locations in the farmland.}
			\label{fig:Infected_Instances_2}
		\end{minipage}
		\hfill
		\begin{minipage}{0.49\textwidth}
			\centering
			\includegraphics[width=0.7\textwidth]{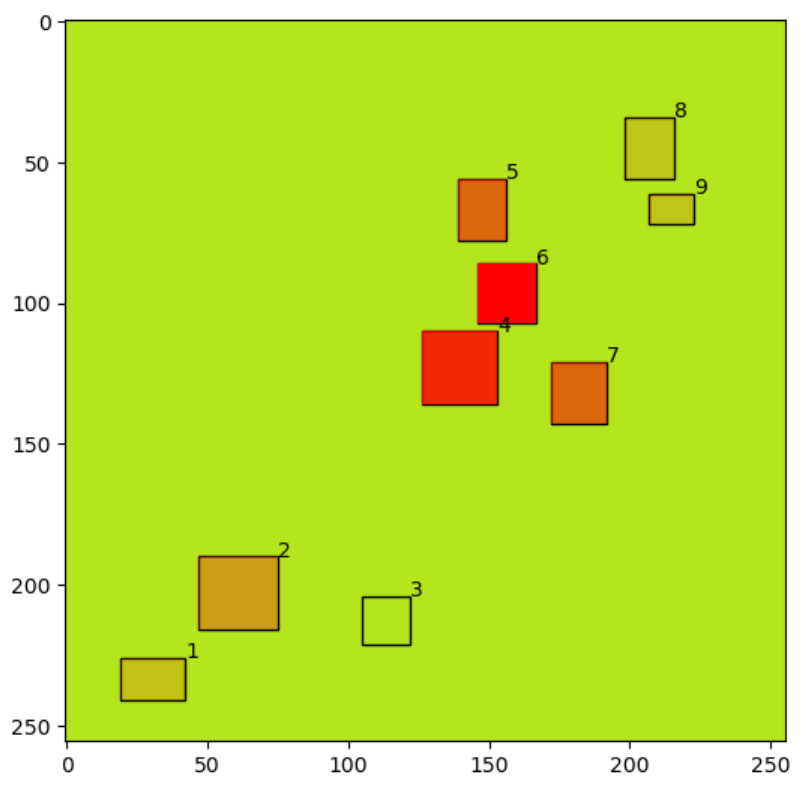}
			\caption{Representation of hotspots}
			\label{fig:Hotspots_Identified_2}
		\end{minipage}
	\end{figure}
\begin{figure}[!htbp]	
			\centering
			\includegraphics[width=0.5\textwidth]{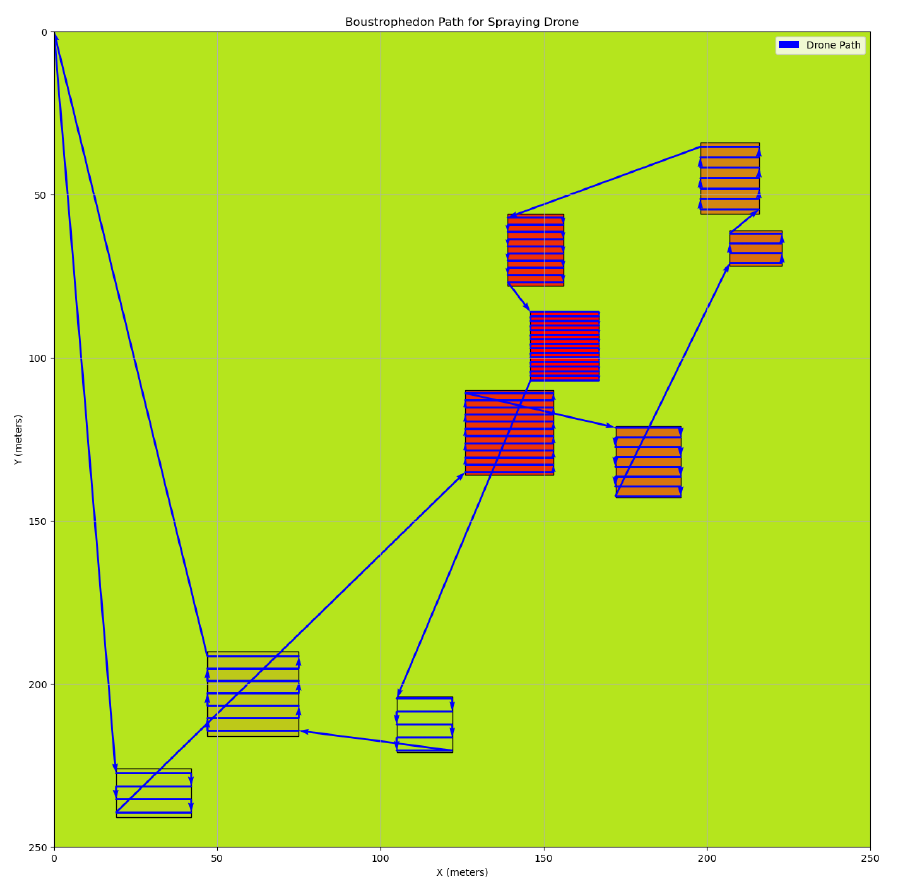}
			\caption{Path computed for spraying drone}
			\label{fig:Drone_Path_Hotspot_2_1}
	\end{figure}

	The results for a different set of locations using variable rate spraying for a drone without in-built rate adapters are presented in Figures \ref{fig:Infected_Instances_2}, \ref{fig:Hotspots_Identified_2}, \ref{fig:Drone_Path_Hotspot_2_1}.
	
	The proposed method can also handle locations specified in GPS coordinates by setting a flag to convert dimensions into meters. The results for the locations shown in Fig \ref{fig:Infected_Instances_2}, submitted as GPS coordinates are presented in Fig \ref{fig:Drone_Path_Hotspot_2_1}.
	

	\begin{figure}[!htbp]
		\centering
		\includegraphics[width=0.7\textwidth]{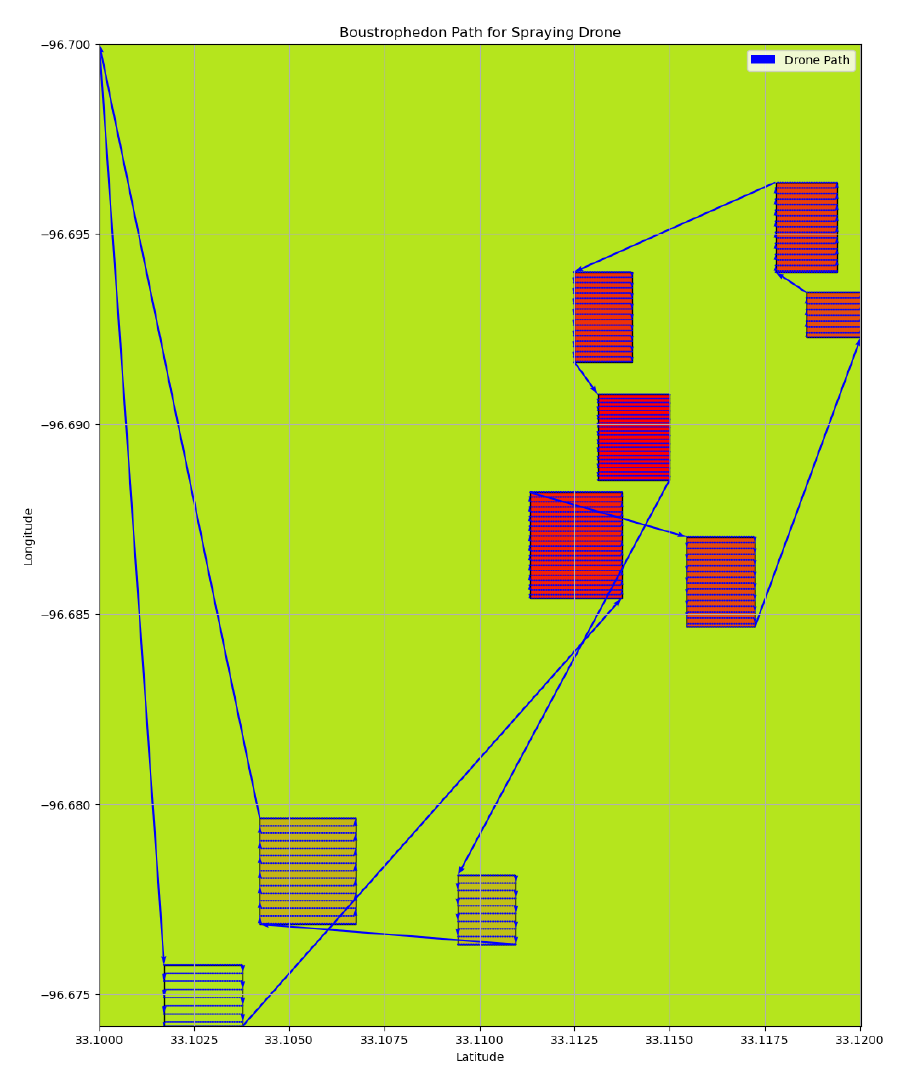}
		\caption{Path computed for spraying drone using GPS coordinates.}
		\label{fig:Drone_Path_Hotspot_GPS}
	\end{figure}
	
	\section{Comparative Perspective With Related Works}
	\label{sec:Comparative_Perspective_With_Related_Works}
	
	SprayCraft stands out by integrating spatial analysis for hotspot detection with a TSP algorithm, optimizing both route length and pesticide usage for variable rate precision spraying. This method addresses limitations in existing works, such as \cite{drones4030058}, \cite{9904967}, \cite{CONESAMUNOZ2016204}, \cite{f12121658},\cite{plessen2024pathplanningspotspraying},\cite{huang2023automatic} which optimize routes but do not support variable rate spraying, and \cite{app8122482}, \cite{7943794}, \cite{8875683},\cite{TEWARI202021} which lacks spatial analysis for hotspot detection. Additionally, though SprayCraft does not support multiple drones as \cite{9994402}, \cite{8264538}, it goes beyond the proposed optimization approaches which do not identify disease hotspots. By combining these advanced features, SprayCraft provides a comprehensive and effective solution for variable rate precision spraying in UAV/drone-based agricultural spraying as briefed in Table \ref{tab:comparison2}.
	
	\begin{table*}[ht]
		\centering
		\caption{Comparison of SprayCraft with Related Works}
		\label{tab:comparison2}
		\begin{tabular}{{p{4cm}p{3cm}p{4cm}p{3cm}}}
			\hline
			\textbf{Work} & \textbf{Route Optimization} & \textbf{Variable Rate Spraying} & \textbf{Hotspot Detection} \\ \hline
			\hline
			\textbf{SprayCraft} & Yes & Yes & Yes \\ \hline
			Wen et al. \cite{app8122482} & No & Yes & No \\ \hline
			Plessen \cite{plessen2024pathplanningspotspraying} & Yes & No & No \\ \hline
			Huang et al. \cite{huang2023automatic} & Yes & No & No \\ \hline
			Tewari et al. \cite{TEWARI202021} & No & Yes & Yes \\ \hline
			Srivastava et al. \cite{drones4030058} & Yes & No & No \\ \hline
			Fang et al. \cite{f12121658} & Yes & No & No \\ \hline
			Xu et al. \cite{9904967} & Yes & No & No \\ \hline
		    {Conesa-Muñoz} et al. \cite{CONESAMUNOZ2016204} & Yes & No & No \\ \hline
			Nolan et al. \cite{7943794} & No & Yes & No \\ \hline
			Muliawan et al. \cite{8875683} & No & Yes & No \\ \hline
			Zheng et al. \cite{9994402} & Yes & No & No \\ \hline
			Lal et al. \cite{8264538} & Yes & No & No \\ \hline
		\end{tabular}
		
	\end{table*}

	\section{Conclusion and Future Work}
	\label{sec:Conclusion}

	 This article, SprayCraft, presented a novel graph-based method for representing diseased locations in farmland. The proposed method effectively identifies hotspots using the graph, computes routes for spraying drones to perform variable rate precision spraying. The same graph can also be utilized to estimate the severity of the damage \cite{10431164}. However, the proposed method is limited to route generation for a single drone and does not account for the impact of wind, which can dynamically deflect droplets from the intended spray area. Additionally, we assume diseased locations to be rectangular, but image segmentation models may identify diseased locations of various shapes.
	 
	 In cases of reduced computing resources, collaborative computing with authentication \cite{Aarella2023} can be implemented to distribute the computational load across multiple devices or systems. Additionally, the results from these computations can be saved on distributed ledgers \cite{Bapatla2022, s22218227}, ensuring data integrity and accessibility for future needs while maintaining trust and transparency in agricultural data management.
	 
	 For future work, developing routes for multiple drones \cite{XU2024142429}, integrating reinforcement learning methods to adjust the drone path based on wind patterns \cite{FAICAL2014393}, and adapting the Boustrophedon Path to the shape of the diseased instances \cite{s21062221} should be considered. These improvements would enhance the efficiency and effectiveness of precision agriculture spraying.


	\balance
	\bibliographystyle{IEEEtran}
	\bibliography{arXiv_2025-XXX_SprayCraft}
	
	\section*{Author Biographies}
	
   	\begin{wrapfigure}{l}{0.17\textwidth}
   		\vspace{-15pt}
   		\includegraphics[width=0.17\textwidth]{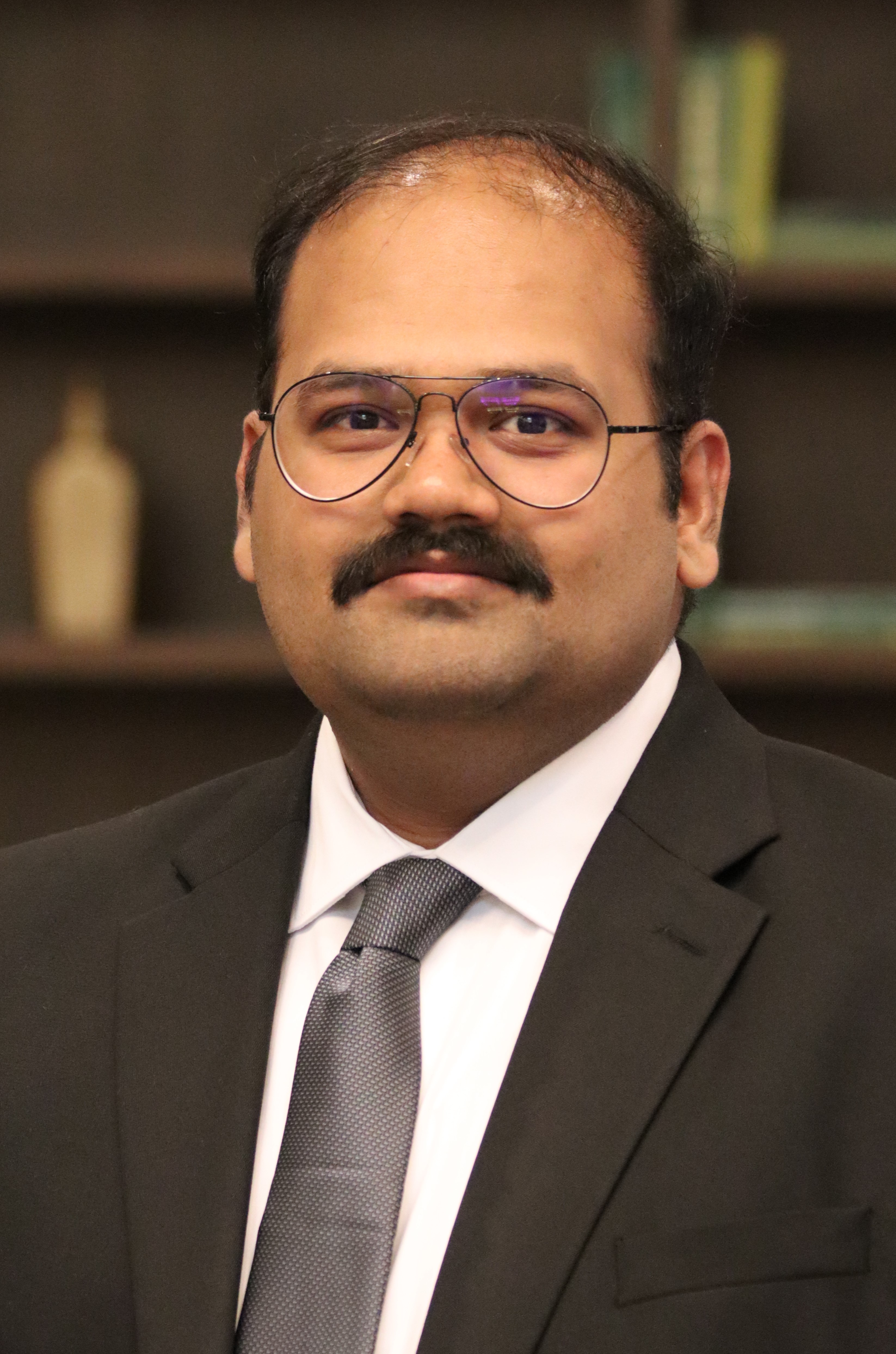}
   		\vspace{-15pt}
   	\end{wrapfigure}
   	\textbf{Kiran K. Kethineni} (Student Member, IEEE) received his B.Tech. degree in Electronics and Communication Engineering from Vignan's University, Guntur, India, in 2016, and his M.S. degree in Computer Engineering from the University of North Texas, Denton, TX, USA, in 2018. He is currently pursuing a Ph.D. in Computer Science and Engineering at the University of North Texas, where he is a member of the Smart Electronics Systems Laboratory research group. His research interests include smart agriculture, as well as exploring the synergy between human and artificial intelligence for applications in the Internet of Things (IoT).
   	
   	\clearpage
   	
   	\begin{wrapfigure}{l}{0.17\textwidth}
   	
   		\includegraphics[width=0.17\textwidth]{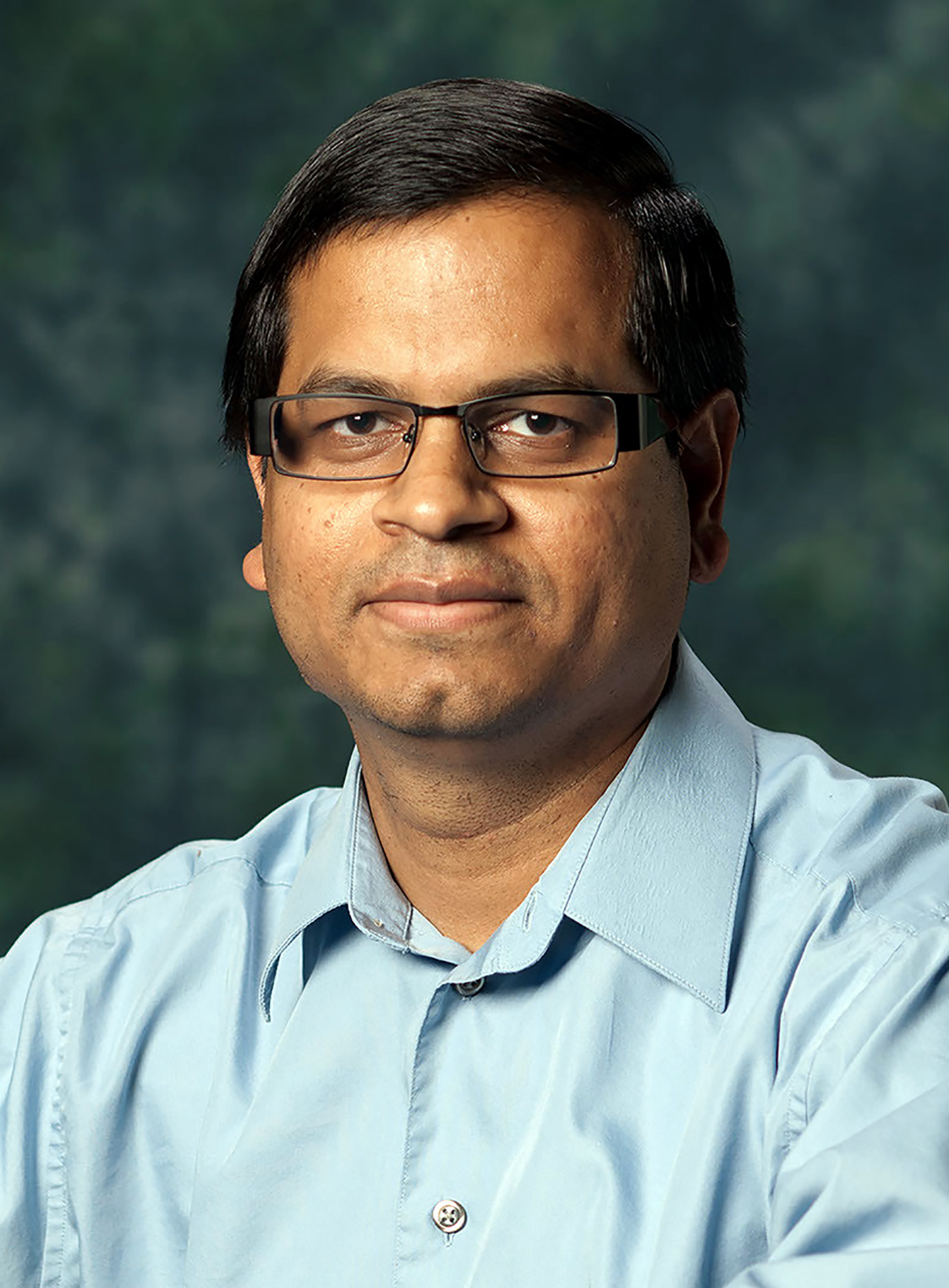}
   		\vspace{-15pt}
   	\end{wrapfigure}	
   	\textbf{Saraju P. Mohanty} (Senior Member, IEEE) received the bachelor’s degree (Honors) in electrical engineering from the Orissa University of Agriculture and Technology, Bhubaneswar, in 1995, the master’s degree in Systems Science and Automation from the Indian Institute of Science, Bengaluru, in 1999, and the Ph.D. degree in Computer Science and Engineering from the University of South Florida, Tampa, in 2003. He is a Professor with the University of North Texas. His research is in ``Smart Electronic Systems’’ which has been funded by National Science Foundations (NSF), Semiconductor Research Corporation (SRC), U.S. Air Force, IUSSTF, and Mission Innovation. He has authored 550 research articles, 5 books, and 10 granted and pending patents. His Google Scholar h-index is 58 and i10-index is 266 with 14,000 citations. He is regarded as a visionary researcher on Smart Cities technology in which his research deals with security and energy aware, and AI/ML-integrated smart components. He introduced the Secure Digital Camera (SDC) in 2004 with built-in security features designed using Hardware Assisted Security (HAS) or Security by Design (SbD) principle. He is widely credited as the designer for the first digital watermarking chip in 2004 and first the low-power digital watermarking chip in 2006. He is a recipient of 19 best paper awards, Fulbright Specialist Award in 2021, IEEE Consumer Electronics Society Outstanding Service Award in 2020, the IEEE-CS-TCVLSI Distinguished Leadership Award in 2018, and the PROSE Award for Best Textbook in Physical Sciences and Mathematics category in 2016. He has delivered 30 keynotes and served on 15 panels at various International Conferences. He has been serving on the editorial board of several peer-reviewed international transactions/journals, including IEEE Transactions on Big Data (TBD), IEEE Transactions on Computer-Aided Design of Integrated Circuits and Systems (TCAD), IEEE Transactions on Consumer Electronics (TCE), and ACM Journal on Emerging Technologies in Computing Systems (JETC). He has been the Editor-in-Chief (EiC) of the IEEE Consumer Electronics Magazine (MCE) during 2016-2021. He served as the Chair of Technical Committee on Very Large Scale Integration (TCVLSI), IEEE Computer Society (IEEE-CS) during 2014-2018 and on the Board of Governors of the IEEE Consumer Electronics Society during 2019-2021. He serves on the steering, organizing, and program committees of several international conferences. He is the steering committee chair/vice-chair for the IEEE International Symposium on Smart Electronic Systems (IEEE-iSES), the IEEE-CS Symposium on VLSI (ISVLSI), and the OITS International Conference on Information Technology (OCIT). He has supervised 3 post-doctoral researchers, 17 Ph.D. dissertations, 28 M.S. theses, and 28 undergraduate projects.\\

   	\begin{wrapfigure}{l}{0.17\textwidth}
   		\vspace{-15pt}
   		\includegraphics[width=0.17\textwidth]{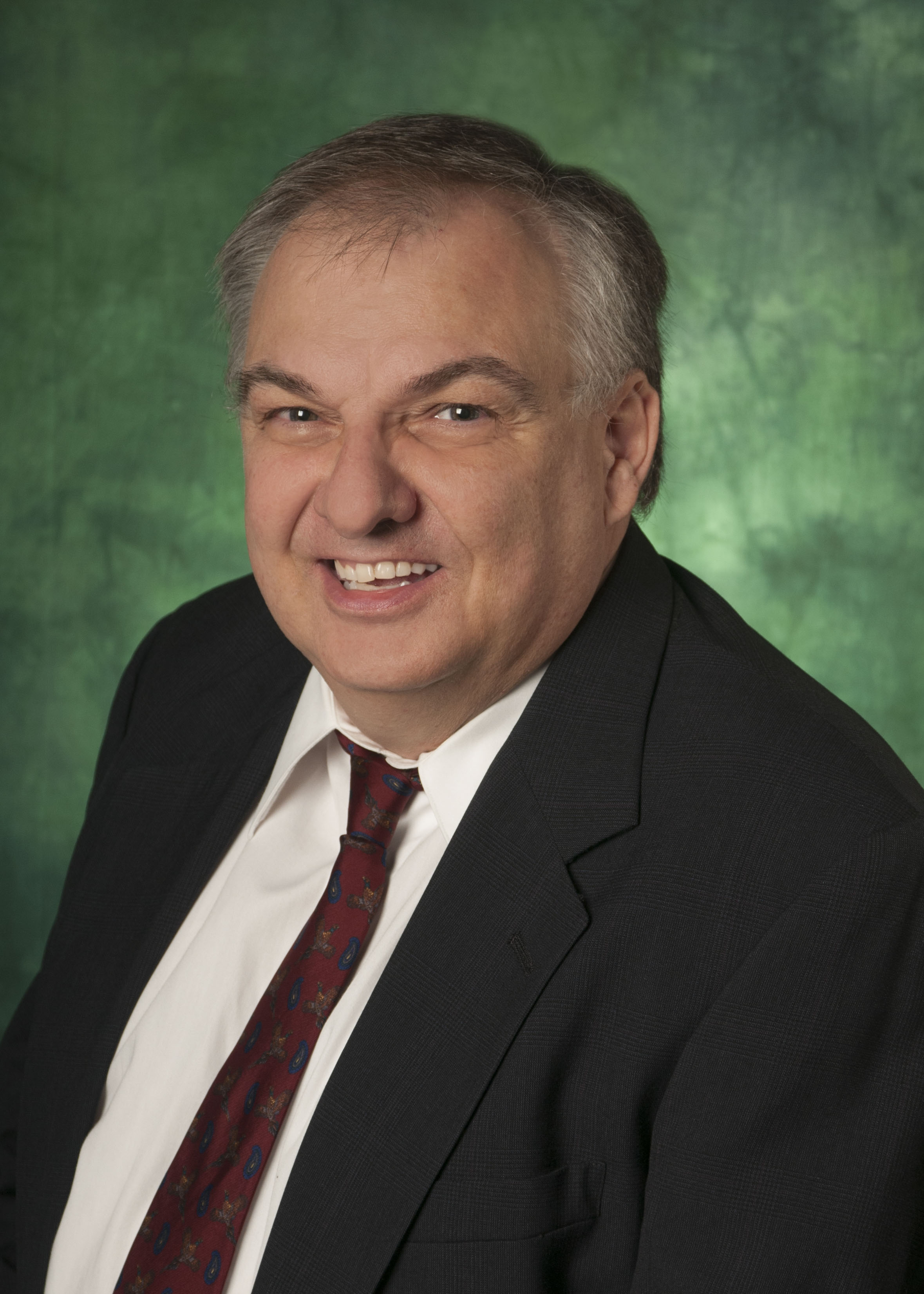}
   		\vspace{-15pt}
   	\end{wrapfigure}	
   	\textbf{Elias Kougianos} received a BSEE from the University of Patras, Greece in 1985 and an MSEE in 1987, an MS in Physics in 1988 and a Ph.D. in EE in 1997, all from Louisiana State University. From 1988 through 1998 he was with Texas Instruments, Inc., in Houston and Dallas, TX. In 1998 he joined Avant! Corp. (now Synopsys) in Phoenix, AZ as a Senior Applications engineer and in 2000 he joined Cadence Design Systems, Inc., in Dallas, TX as a Senior Architect in Analog/Mixed-Signal Custom IC design. He has been at UNT since 2004. He is a Professor in the Department of Electrical Engineering, at the University of North Texas (UNT), Denton, TX. His research interests are in the area of Analog/Mixed-Signal/RF IC design and simulation and in the development of VLSI architectures for multimedia applications. He is an author of over 200 peer-reviewed journal and conference publications.\\

   	\begin{wrapfigure}{l}{0.17\textwidth}
   		\vspace{-15pt}
   		\includegraphics[width=0.17\textwidth]{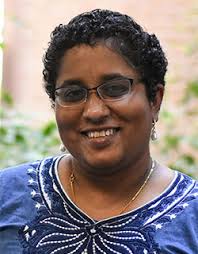}
   		\vspace{-15pt}
   	\end{wrapfigure}
   	\textbf{Sanjukta Bhowmick} is an Associate Professor in the Computer Science and Engineering department at the University of North Texas. She obtained her Ph.D. from the Pennsylvania State University. She had a joint postdoc at Columbia University and Argonne National Lab. Her current research is on understanding change in complex network analysis, with a focus on developing scalable algorithms for large dynamic networks and developing uncertainty quantification metrics for network analysis. Dr. Bhowmick has served in the leadership roles in several conferences including Supercomputing, ISC, IPDPS, and HiPC. Currently, she is the co-chair of the IEEE TCHPC (Technical Consortium on High Performance Computing) Education Outreach Initiative.
   	
   	\clearpage
	
	\begin{wrapfigure}{l}{0.17\textwidth}
		
		\includegraphics[width=0.17\textwidth]{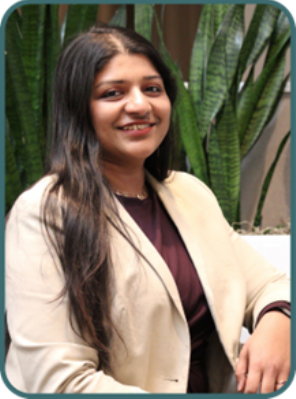}
		\vspace{-15pt}
	\end{wrapfigure}
	\textbf{Laavanya Rachakonda} is an Assistant Professor in the Department of Computer Science in the College of Science and Engineering at the University of North Carolina Wilmington, Wilmington, NC. She earned her Bachelor of Technology (B. Tech) in Electronics and Communication from Jawaharlal Nehru Technological University (JNTU), Hyderabad, India, Master of Sciences (M.S) in Computer Engineering, and Doctor of Philosophy (Ph.D.) in Computer Science and Engineering from University of North Texas. During her graduate studies, she was part of the Smart Electronics Systems Laboratory (SESL) research group at Computer Science and Engineering at the University of North Texas, Denton, TX. Her research interests include smart healthcare applications using artificial intelligence, deep learning approaches, and application-specific architectures for consumer electronic systems based on the IoT. She has 3 peer-reviewed journals published, 13 peer-reviewed conference publications, 2 filed patents, and 1 patent disclosure. She has delivered 15 talks (online and offline) at various IEEE-hosted conferences. She has won 20 honors and awards and has monitored 6 undergraduate and TAMS students. Her biography, research, education, and outreach activities are available at www.laavanyarachakonda.com.

\end{document}